\documentclass[11pt,a4paper]{scrartcl}
\usepackage{hyperref,amsmath,amsthm,amsfonts,amstext, amssymb, tikz, xspace, cite, colonequals, hyperref, enumitem, array, rotating, wasysym,multirow}
\usetikzlibrary{shapes.geometric}
\usetikzlibrary{decorations.pathreplacing}
\usepackage[linesnumbered,boxruled,lined,noend]{algorithm2e}

\begin{document}

\title{A Fast Algorithm for Permutation Pattern Matching Based on Alternating Runs
\thanks{A preliminary version of this paper~\cite{DBLP:conf/swat/BrunerL12} appeared in the proceedings of the 13th Scandinavian Symposium and Workshops on Algorithm Theory, SWAT 2012.}
}

\author{Marie-Louise Bruner\\Insitute of Discrete Mathematics and Geometry\\Vienna University of Technology, Austria\\
              \texttt{marie-louise.bruner@tuwien.ac.at} \and Martin Lackner\\Insitute of Information Systems\\Vienna University of Technology, Austria\\\texttt{lackner@dbai.tuwien.ac.at}}

\date{}

  \newtheorem{theorem}{Theorem}[section]
  \newtheorem{corollary}[theorem]{Corollary}
  \newtheorem{claim}[theorem]{Claim}
  \newtheorem{openquestion}[theorem]{Open Question}
  \newtheorem{pausetoponder}[theorem]{Pause to Ponder}
  \newtheorem{goal}[theorem]{Goal}
  \newtheorem{lemma}[theorem]{Lemma}
  \newtheorem{observation}[theorem]{Observation}
  \newtheorem{remark}[theorem]{Remark}
  \newtheorem{fact}[theorem]{Fact}
  \newtheorem{proposition}[theorem]{Proposition}
  \newtheorem{definition}[theorem]{Definition}
  \newtheorem{example}[theorem]{Example}
  \newtheorem{construction}[theorem]{Construction}

\newcommand{\cprob}[3]{
    \begin{center}
      \fbox{
        \parbox{0.9\textwidth}{
          #1\smallskip\\
          \begin{tabular}{rp{0.72\textwidth}}
            \textit{Instance:\ } & #2\\
            \textit{Question:\ } & #3
          \end{tabular}
        }
      }
    \end{center}
}

\newcommand{\pprob}[4]{
    \begin{center}
      \fbox{
        \parbox{0.9\textwidth}{
          #1\smallskip\\
          \begin{tabular}{rp{0.70\textwidth}}
            \textit{Instance:\ } & #2\\
            \textit{Parameter:\ } & #3\\
            \textit{Question:\ } & #4
          \end{tabular}
        }
      }
    \end{center}
}

\newcommand{\unt}{\underline}
\newcommand{\pe}{permutation}
\newcommand{\mul}{multiset}
\newcommand{\pea}{permutation }
\newcommand{\mula}{multiset }

\newcommand{\vale}{\mathsf{vale}}
\newcommand{\run}{\mathsf{run}}
\newcommand{\pre}{\mathsf{pre}}
\newcommand{\ri}{\mathsf{ri}}
\newcommand{\vi}{\mathsf{vi}}
\newcommand{\lrun}{\mathsf{lrun}}
\newcommand{\jump}{\mathsf{jump}}
\newcommand{\asc}{\mathsf{asc}}
\newcommand{\des}{\mathsf{des}}
\newcommand{\dis}{\mathsf{dis}}
\newcommand{\inv}{\mathsf{inv}}
\newcommand{\sinv}{\mathsf{sinv}}
\newcommand{\maxisc}{\mathsf{maxisc}}
\newcommand{\maxls}{\mathsf{maxls}}
\newcommand{\minisc}{\mathsf{minisc}}
\newcommand{\minls}{\mathsf{minls}}
\newcommand{\svc}{\mathsf{svc}}

\newcommand {\cA} {{\mathcal A}}
\newcommand {\cC} {{\mathcal C}}
\newcommand {\cI} {{\mathcal I}}
\newcommand {\cS} {{\mathcal S}}
\newcommand {\cX} {{\mathcal X}}
\newcommand {\cY} {{\mathcal Y}}
\newcommand {\cM} {{\mathcal M}}
\newcommand {\cZ} {{\mathcal Z}}
\newcommand {\cL} {{\mathcal L}}
\newcommand {\cT} {{\mathcal T}}
\newcommand {\cN} {{\mathcal N}}
\newcommand {\cB} {{\mathcal B}}
\newcommand {\cP} {{\mathcal P}}
\newcommand {\cV} {{\mathcal V}}
\newcommand {\cF} {{\mathcal F}}

\newcommand {\bigO} {{\mathcal O}}

\newcommand {\N} {{\mathbb N}}

\newcommand{\ccfont}[1]{\textsf{#1}}
\newcommand{\probfont}[1]{\textsc{#1}}

\newcommand{\ppm}{\probfont{PPM}\xspace}
\newcommand{\gppm}{\probfont{VPPM}\xspace}
\newcommand{\vppm}{\probfont{VPPM}\xspace}

\newcommand{\fpt}{\textnormal{\ccfont{FPT}}\xspace}
\newcommand{\xp}{\textnormal{\ccfont{XP}}\xspace}
\newcommand{\w}[1]{\ifmmode{\textnormal{\ccfont{W[#1]}}}\else{\textnormal{\ccfont{W[#1]}}}\fi}
\newcommand{\varMC}[1]{\probfont{var-MC}[\ensuremath{#1}]}
\newcommand{\MC}[1]{\probfont{MC}[\ensuremath{#1}]}

\newcommand{\var}[1]{\ensuremath{\textsl{var}\,(#1)}}
\newcommand{\lit}[1]{\ensuremath{\textsl{lit}\,(#1)}}
\newcommand{\card}[1]{\ensuremath{|#1|}}
\newcommand{\size}[1]{\ensuremath{\left\lVert #1 \right\rVert}}
\newcommand{\ra}{\rightarrow}
\newcommand{\true}{\textsl{true}\xspace}
\newcommand{\false}{\textsl{false}\xspace}

\newcommand{\Ptime}{\textnormal{\ccfont{P}}\xspace}
\newcommand{\NP}{\ccfont{NP}\xspace}
\newcommand{\phd}[1]{\ensuremath{\mathsf{\Delta}_{#1}^\Ptime}}
\newcommand{\phs}[1]{\ensuremath{\mathsf{\Sigma}_{#1}^\Ptime}}
\newcommand{\php}[1]{\ensuremath{\mathsf{\Pi}_{#1}^\Ptime}}

\makeatletter
\def\underbracket{%
\@ifnextchar[{\@underbracket}{\@underbracket [\@bracketheight]}%
}
\def\@underbracket[#1]{%
\@ifnextchar[{\@under@bracket[#1]}{\@under@bracket[#1][0.4em]}%
}
\def\@under@bracket[#1][#2]#3{
\mathop{\vtop{\m@th \ialign {##\crcr $\hfil \displaystyle {#3}\hfil $%
\crcr \noalign {\kern 3\p@ \nointerlineskip }\upbracketfill {#1}{#2}
\crcr \noalign {\kern 3\p@ }}}}\limits}
\def\upbracketfill#1#2{$\m@th \setbox \z@ \hbox {$\braceld$}
\edef\@bracketheight{\the\ht\z@}\bracketend{#1}{#2}
\leaders \vrule \@height #1 \@depth \z@ \hfill
\leaders \vrule \@height #1 \@depth \z@ \hfill \bracketend{#1}{#2}$}
\def\bracketend#1#2{\vrule height #2 width #1\relax}
\makeatother

\newcommand{\und}[1]{\underbracket[0.5pt]{#1}}

\newcommand{\exend}{\ifmmode\hbox{$\dashv$}\else{\unskip\nobreak\hfil\penalty50\hskip1em\null\nobreak\hfil\hbox{$\dashv$}
\parfillskip=0pt\finalhyphendemerits=0\endgraf}\fi}

\newcommand{\defend}{\ifmmode\hbox{$\dashv$}\else{\unskip\nobreak\hfil
\penalty50\hskip1em\null\nobreak\hfil\hbox{$\dashv$}
\parfillskip=0pt\finalhyphendemerits=0\endgraf}\fi}

\newcommand\xqed[1]{%
  \leavevmode\unskip\penalty9999 \hbox{}\nobreak\hfill
  \quad\hbox{#1}}
\newcommand\demo{\xqed{$\dashv$}}

\newcommand\RotText[1]{\rotatebox{90}{\parbox{3cm}{\centering#1}\ }}

\newcommand\nop[1]{}

\def\getchar{\let\char= }
\def\perm#1{\permutationloop #1\nil}
\def\permutationloop{\afterassignment\gaporsqueeze\getchar}
\def\gaporsqueeze{%
  \ifx\char\nil%
     \hspace{-\s}\let\next=\relax%
  \else%
    \ifx,\char%
      \hspace{-\s}\hspace{\g}\let\next=\permutationloop%
    \else%
      \char\hspace{\s}\let\next=\permutationloop%
    \fi%
  \fi%
  \next%
}
\def\g{0.06cm}
\def\s{-0.05cm}
\newcommand{\Tex}{\perm{1,8,12,4,7,11,6,3,2,9,5,10}}
\renewcommand{\vec}[1]{\bar{#1}}

\maketitle

\begin{abstract}
\quad The \NP-complete \textsc{Permutation Pattern Matching} problem asks whether a $k$-permutation $P$ is contained in a $n$-permutation $T$ as a pattern.
This is the case if there exists an order-preserving embedding of $P$ into $T$.
In this paper, we present a fixed-parameter algorithm solving this problem with a worst-case runtime of $\bigO(1.79^{\run(T)}\cdot n\cdot k)$, where $\run(T)$ denotes the number of alternating runs of $T$.
This algorithm is particularly well-suited for instances where $T$ has few runs, i.e., few ups and downs.
Moreover, since $\run(T)<n$, this can be seen as a $\bigO(1.79^{n}\cdot n\cdot k)$ algorithm which is the first to beat the exponential $2^n$ runtime of brute-force search.
Furthermore, we prove that under standard complexity theoretic assumptions such a fixed-parameter tractability result is not possible for $\run(P)$.
\end{abstract}

\section{Introduction}
The concept of pattern matching in permutations arose in the late 1960ies. It was in an exercise of his \textit{Fundamental algorithms} \cite{DBLP:books/aw/Knuth68} that Knuth asked which permutations could be sorted using a single stack.
The answer is simple: These are exactly the permutations that do not contain the pattern $231$.
By containing a certain pattern the following is meant:
The permutation $\pi=53142$ (written in one-line representation) contains the pattern $231$, since the subsequence $342$ of $\pi$ is order-isomorphic to $231$.
We call the function $\{2\mapsto 3, 3\mapsto 4, 1\mapsto 2\}$ a matching of $231$ into $\pi$.
On the other hand, $\pi$ does not contain the pattern $123$ since it contains no increasing subsequence of length three. 
Since 1985, when the first systematic study of \textit{Restricted Permutations}~\cite{simion1985restricted} was published by Simion and Schmidt, the area of permutation patterns has become a rapidly growing field of discrete mathematics, more specifically of combinatorics~\cite{bona_combinatorics_2004,kitaev2011patterns,vatterPP}.
Many applications of permutation patterns have been discovered:
their relation to stack and deque sorting, genome sequences in computational biology, statistical mechanics and in general their numerous connections to other combinatorial objects~\cite{kitaev2011patterns}.

This paper takes the viewpoint of computational complexity.
Computational aspects of permutation patterns, in particular the analysis of the \textsc{Permutation Pattern Matching} (\ppm) problem, have received far less attention than enumerative questions until now. The \ppm problem is defined as follows:

\cprob {\probfont{Permutation Pattern Matching} (\ppm)}
{A \pea $T$ (the text) of length $n$ and a permutation $P$ (the pattern) of length $k\leq n$.}
{Is there a matching of $P$ into $T$?}
Bose, Buss and Lubiw \cite{Bose1998277} showed that \ppm is in general \NP-complete.
The trivial brute-force algorithm checking every subsequence of length $k$ of $T$ has a runtime of $\bigO(2^n\cdot n)$.
So far, no algorithm has been discovered that improves the exponential runtime to $c^n$ for some constant $c<2$.
Improving exponential time algorithms is a major topic in algorithmics, as witnessed by the monograph of Fomin and Kratsch~\cite{fomin-book}.

In this paper we tackle the problem of solving \ppm faster than $\bigO(2^n\cdot n)$ for arbitrary $P$ and $T$. 
We achieve this by exploiting the decomposition of permutations into alternating runs.
As an example, the permutation $\pi=53142$ has three alternating runs: $531$ (down), $4$ (up) and $2$ (down).
We denote this number of ups and downs in a permutation $\pi$ by $\run(\pi)$.
Alternating runs are a fundamental permutation statistic and were studied already in the late 19th century by Andr\'{e}~\cite{andre1884}.
Despite the importance of alternating runs within the study of permutations, the connection to \ppm has so far not been explored.

\paragraph*{Contributions.}\ 
The contributions of this paper are the following:
\begin{itemize}
\item Our main result is a fixed-parameter tractable algorithm for \ppm with a runtime of $\bigO(1.79^{\run(T)}\cdot n\cdot k)$.
Since the combinatorial explosion is confined to $\run(T)$, this algorithm performs especially well when $T$ has few alternating runs.
\item Since $\run(T)\leq n$, this algorithm also solves \ppm in time $\bigO(1.79^n\cdot n\cdot k)$. 
This is a major improvement over the brute-force algorithm with a runtime of $\mathcal{O}(2^n\cdot n)$.
\item 
Since the the number of runs in a random permutation is unlikely to be $n$, one can expect an even smaller constant than $1.79$ on average.
Indeed, we prove that the expected runtime of our algorithm is in $\bigO(1.52^n\cdot n\cdot k)$. 
\item We also show that an algorithm by Ahal and Rabinovich~\cite{DBLP:journals/siamdm/AhalR08} has a runtime of $\bigO(n^{1+\run(P)})$.
This is achieved by proving that the pathwidth of a certain graph generated by a permutation is bounded by the number of alternating runs of this permutation.
\item Finally, we prove that -- under standard complexity theoretic assumptions -- no fixed-parameter algorithm exists with respect to $\run(P)$, i.e., no algorithm with runtime $\bigO(c^{\run(P)}\cdot \mathit{poly}(n))$ for some constant $c$ may be hoped for.
Thus, the runtime of the aforementioned $\bigO(n^{1+\run(P)})$ algorithm cannot be substantially improved.
\end{itemize}

\paragraph*{Related work.}\ 
The most relevant paper is the recent break-through result by Guillemot and Marx~\cite{guillemotmarx2013ppmfpt} showing that \ppm is FPT with respect to the length of the pattern.
Their algorithm has a runtime of $2^{\bigO(k^2\cdot log k)}\cdot n$.
This FPT result is anteceeded by algorithms with a runtime of $\bigO(n^{1+2k/3}\cdot\log n)$~\cite{springer:AlbertAAH01} and $\bigO(n^{0.47k+o(k)})$~\cite{DBLP:journals/siamdm/AhalR08}.
\ppm has also been studied for more general types of patterns.
For these more general cases, it has been shown that FPT results with parameter $k$ are not possible~\cite{landscape}.

Although \ppm is \NP-complete in general, there are polynomial time algorithms if only certain permutations are allowed as patterns.
The most important example are \textit{separable} permutations: these are permutations that contain neither $3142$ nor $2413$.
If the pattern is seperable, \ppm can be solved in polynomial time~\cite{Bose1998277,DBLP:journals/ipl/Ibarra97,springer:AlbertAAH01,DBLP:journals/dam/YugandharS05}.
In case $P$ is the identity $\perm{1,2,\ldots,k}$, \ppm consists of looking for an increasing subsequence of length $k$ in the text $T$ -- this is a special case of the \probfont{Longest Increasing Subsequence} problem.
This problem can be solved in $\bigO(n \log n)$-time for sequences in general \cite{schensted1987longest} and in $\mathcal{O}(n \log \log n)$-time for permutations \cite{DBLP:journals/ipl/ChangW92,maekinen2001longest}.
An $\bigO(k^2n^6)$-time algorithm is presented in \cite{springer:GuillemotV09} for the case that both the text and the pattern that do not contain $321$.
If the pattern is required to be matched to consecutive elements in the text, a $\bigO(n+k)$ algorithm has been found~\cite{Kubica2013430}.
A similar result has been found independently in~\cite{DBLP:journals/corr/abs-1302-4064}.
This work has been extended to the cases where some mismatches are tolerated~\cite{DBLP:journals/corr/GawrychowskiU13}; also suffix trees have recently been generalized to be applicable in this setting~\cite{DBLP:conf/spire/CrochemoreIKKLPRRW13}.

The related \probfont{Longest Common Pattern} problem is to find a longest common pattern between two permutations $T_1$ and $T_2$, i.e., a pattern $P$ of maximal length that can be matched both into $T_1$ and $T_2$. This problem is a generalization of \ppm since determining whether $T_1$ is the longest common pattern between $T_1$ and $T_2$ is equivalent to checking whether $T_2$ contains $T_1$ as a pattern. In \cite{bouvel_longest_2006} a polynomial time algorithm for the \probfont{Longest Common Pattern} problem is presented for the case that one of the two permutations $T_1$ and $T_2$ is separable. A generalization of this problem, the so called \probfont{Longest Common $\mathcal{C}$-Pattern} problem was introduced in~\cite{DBLP:conf/cpm/BouvelRV07}. This problem consists of finding the longest common pattern among several permutations belonging to a class $\mathcal{C}$ of permutations. For the case that $\mathcal{C}$ is the class of all separable permutations and that the number of input permutations is fixed, the problem was shown to be polynomial time solvable~\cite{DBLP:conf/cpm/BouvelRV07}.

For a class of permutations $X$, the  \probfont{Longest $X$-Subsequence} (LXS) problem is to identify in a given permutation $T$ its longest subsequence that is isomorphic to a permutation of $X$. 
Polynomial time algorithms for many classes $X$ exist, but in general \probfont{LXS} is \NP-hard~\cite{albert2003longest}.

\paragraph*{Organization of the paper.}\ 
Section~\ref{sec:prel} contains essential definitions for permutations and parameterized complexity theory.
The main section, Section~\ref{sec:alg}, describes the algorithm and is divided into the following subsections.
Section~\ref{subsec:matchingfunctions} introduces matching functions.
Section~\ref{subsec:algdesc} describes the alternating run algorithm in detail.
Section~\ref{subsec:correctness} contains proof details necessary to verify the correctness of the alternating run algorithm.
Section~\ref{subsec:runtime} proves the corresponding runtime bounds.
Our results concerning the parameter $\run(P)$ can be found in Section~\ref{sec:hard}.
We conclude with future research directions in Section~\ref{sec:future}.

\section{Preliminaries}
\label{sec:prel}
\subsection{Permutations}

For any $m\in\mathbb{N}$, let $[m]$ denote the set $\{1,\dots,m\}$.
For $m\leq n\in\mathbb{N}$, $[m,n]$ denotes the set $\{m,m+1,\ldots,n\}$.
A permutation is a bijective function from a finite set onto itself.
An $m$-permutation is a permutation from $[m]$ to $[m]$.
An $m$-permutation $\pi$ can be seen as the sequence $\pi(1), \pi(2), \ldots, \pi(m)$.
We distinguish between elements of permutations ($\pi(1), \pi(2), \ldots$) and their positions ($1,2,\ldots$).
Viewing permutations as sequences allows us to speak of \emph{subsequences} of a permutation.
We speak of a \emph{contiguous subsequence} of $\pi$ if the sequence consists of contiguous elements in $\pi$, i.e., if the corresponding positions form an interval.
Given a set $S\subseteq[m]$, we write $\pi \vert_S$ to denote the subsequence of $\pi$ consisting exactly of the elements of $S$.
\begin{definition}
Let $P$ (the pattern) be a $k$-permutation. We say that an $n$-permutation $T$ (the text) \emph{contains $P$ as a pattern} or that \emph{$P$ can be matched into $T$} if we can find a subsequence of $T$ that is order-isomorphic to $P$. 
Matching $P$ into $T$ thus consists in finding a monotonically increasing map $M: [k] \rightarrow [n]$ so that the sequence $M(P)$, defined as  $\big(M(P(1)),M(P(2)),\ldots,M(P(k))\big)$, is a subsequence of $T$.
Such a function $M$ is then called a \emph{matching}.
Note that $M$ maps elements of $P$ to elements of $T$.
\label{Def.pattern.avoid}%
\end{definition}

\begin{example}
We will use the text permutation $T_{\mathit{ex}}=\Tex$ and the pattern permutation $P_{\mathit{ex}}=\perm{2,3,1,4}$ as a running example throughout the paper.
A graphical representation can be found in Figure~\ref{fig:ex_F} on page~\pageref{fig:ex_F}.
The pattern $P_{\mathit{ex}}$ can be matched into $T_{\mathit{ex}}$ as witnessed by the subsequence $\perm{4,6,2,9}$.
\demo
\end{example}

Every $[m]$-permutation $\pi$ defines a total order $\prec _{\pi}$ on $[m]$.
We write $u \prec_{\pi} v$ if $\pi ^{-1}(u) < \pi ^{-1}(v)$, i.e., the element $u$ stands to the left of the element $v$ in $\pi$.
We write $u \preceq_{\pi} v$ if either $u \prec_{\pi} v$ or $u=v$.
We say $u$ is left (right) of $v$ if $u \prec _{\pi} v$ ($v \prec _{\pi} u$).

We discern two types of local extrema in permutations: valleys and peaks.
A \emph{valley} of a permutation $\pi$ is an element $\pi(i)$ for which it holds that $\pi(i-1) > \pi(i)$ and $\pi(i) < \pi(i+1)$.
If $\pi(i-1)$ or $\pi(i+1)$ is not defined, we still speak of valleys.
Similarly, a \emph{peak} denotes an element $\pi(i)$ for which it holds that $\pi(i-1) < \pi(i)$ and $\pi(i) > \pi(i+1)$. 

Valleys and peaks partition a permutation into contiguous monotone subsequences, so-called \emph{(alternating) runs}.
The first run of a given permutation starts with its first element (which is also the first local extremum) and ends with the second local extremum.
The second run starts with the following element and ends with the third local extremum.
Continuing in this way, every element of the permutation belongs to exactly one alternating run.
Observe that every alternating run is either increasing or decreasing.
We therefore distinguish between \emph{runs up} and \emph{runs down}.
Note that runs up always end with peaks and runs down always end with valleys.
The parameter $\run(\pi)$ counts the number of alternating runs in $\pi$.
Hence, $\run(\pi)+1$ equals the number of local extrema in $\pi$.
These definitions can be analogously extended to subsequences of permutations.

\begin{example}
In the permutation $T_\mathit{Ex}=\Tex$ the valleys are $1$, $4$, $2$ and $5$ and the peaks are $\perm{12}$, $\perm{11}$, $9$ and $\perm{10}$. A decomposition into alternating runs is given by: $\perm{1,8,12}|4|\perm{7,11}|\perm{6,3,2}|9|5|\perm{10}$.
\demo
\end{example}

The following two functions only concern the pattern $P$ and not arbitrary permutations.
Let $u \in [k]$.
The \emph{run predecessor} $\pre(u)$ denotes the largest element smaller than $u$ that is contained in the same run as $u$ in $P$ (if such an element exists).
Moreover, the \emph{run index} function $\ri$ is defined as follows: $\ri(u)=i$ if $u$ is contained in the $i$-th run in $P$.

\subsection{Parameterized complexity theory}

We give the relevant definitions of parameterized complexity theory. 
In contrast to classical complexity theory, a parameterized complexity analysis studies the runtime of an algorithm with respect to an additional parameter and not just the input size~$\card{I}$. Therefore, every parameterized problem is considered as a subset of $\Sigma^*\times \N$, where $\Sigma$ is the input alphabet. 
An instance of a parameterized problem consequently consists of an input string together with a positive integer $p$, the parameter. 
\begin{definition}
A parameterized problem is \emph{fixed-parameter tractable} (or in \fpt) if there is a computable function $f$ and an integer $c$ such that 
there is an algorithm solving an instance $(I,p)$ in time $\bigO(f(p)\cdot \card{I}^c)$.
\end{definition}
The algorithm itself is also called fixed-parameter tractable (FPT).

A central concept in parameterized complexity theory are \emph{fixed-parameter tractable reductions}, which allow for a parameterized hardness theory.
\begin{definition}
Let $L_1,L_2\subseteq \Sigma^*\times \N$ be two parameterized problems.
An \emph{FPT-reduction} from $L_1$ to $L_2$ is a mapping $G : \Sigma^*\times \N \ra \Sigma^*\times \N$ such that
\begin{itemize}
\item $(I, p) \in L_1$ if and only if $G(I, p) \in L_2$.
\item $G$ is computable by an FPT-algorithm.
\item There is a computable function $H$ such that for $G(I, p) = (I' , p')$, $p' \leq H(p)$ holds.
\end{itemize}
\end{definition}

Besides the class \fpt, other important complexity classes in the framework of parameterized complexity are $\w{1}\subseteq\w{2}\subseteq\ldots$, the \textnormal{\ccfont{W}}-hierarchy.
For our purpose, only the class \w{1} is relevant.
It is conjectured (and widely believed) that $\w{1}\neq \fpt$. 
Therefore, showing \w{1}-hardness can be considered as evidence that the problem is not fixed-parameter tractable.

\begin{definition}
The class $\w{1}$ is defined as the class of all problems that are FPT-reducible to the following problem.
\end{definition}
\pprob
{\probfont{Turing Machine Acceptance}}
{A nondeterministic Turing machine with its transition table,
an input word $x$ and a positive integer $k$.}
{$k$}
{Does the Turing machine accept the input $x$ in at most $k$ steps?}

\begin{definition}
A parameterized problem is in \xp if it can be solved in time $\bigO(\card{I}^{f(k)})$ where $f$ is a computable function.
\end{definition}

\noindent All the aforementioned classes are closed under FPT-reductions. 
The following relations between these complexity classes are known:
\[
\fpt\subseteq\w{1}\subseteq\w{2}\subseteq\ldots\subseteq\xp
\]

Further details can be found for example in the monographs by Downey and Fellows~\cite{DowneyF99Book,fellows-neu}, Niedermeier~\cite{niedermeier2006invitation} and Flum and Grohe~\cite{FlumG2006parameterized}.

\begin{remark}
For our runtime considerations we assume the random access machine (RAM) model~\cite{machinemodels}.
In addition, we assume that elementary operations (addition, subtraction, multiplication, modulo) require constant time.
\end{remark}

\paragraph{Notation.}\ 
We use capital letters for functions, sets and lists and lower case letters for natural numbers.
The letters $i$ and $j$ are exclusively used to denote positions or indices such as positions in permutations or indices in lists.
We use greek letters only in three cases:
$\pi$ is an arbitrary permutation and $\kappa\in[k]$ as well as $\nu\in[n]$ are the main variables in the algorithm.
In this paper, tuples always have length $\run(P)$ and are denoted with bars, e.g., $\vec{x}$.
The elements of $\vec{x}$ are $x_1,\ldots,x_{\run(P)}$.

\section{The alternating run algorithm}
\label{sec:alg}
We start with an outline of the alternating run algorithm. 
Its description consists of two parts.
In Part 1 we introduce so-called \textit{matching functions}.
These functions map runs in $P$ to sequences of adjacent runs in $T$.
The intention behind matching functions is to restrict the search space to certain subsequences of length $k$, namely to those where all elements in a run in $P$ are mapped to elements in the corresponding sequences of runs in $T$.
In Part~2 a dynamic programming algorithm is described.
It checks for every matching function whether it is possible to find a compatible matching.
This is done by finding a small set of representative elements to which the element $1$ can be mapped to, then -- for a given choice for $1$ -- finding representative elements for $2$, and so on.
\begin{theorem}
\label{thm-ppm-altruns}
The alternating run algorithm solves \probfont{Permutation Pattern Matching} in time $\bigO(1.79^{\run (T)}\cdot n\cdot k)$. Therefore, \probfont{Permutation Pattern Matching} parameterized by $\run (T)$ is in \fpt.
\end{theorem}

Since $\run(T)<n$, we obtain as an immediate consequence:

\begin{corollary}
\label{thm-ppm-n}
The alternating run algorithm solves the \probfont{Permutation Pattern Matching} problem in time $\bigO(1.79^{n}\cdot n\cdot k)$.
\end{corollary}

\subsection{Matching functions.}
\label{subsec:matchingfunctions}

We introduce the concept of matching functions.
These are functions from the interval $[\run(P)]$ to sequences of adjacent runs in $T$.
For a given matching function $F$ the search space in $T$ is restricted to matchings where an element $\kappa$ contained in the $i$-th run in $P$ is matched to an element in $F(i)$.
As we will see later on in Lemma~\ref{lem:compatible}, this restriction of the search space does not influence whether a matching can be found or not: if a matching exists, a corresponding matching function can be found.
In addition, Lemma~\ref{lem:count functions F} will show that it is possible to iterate over all matching functions in FPT time.
Thus, our algorithm verifies for all matching functions whether a compatible matching exists.

Let us now give a formal definition of matching functions.

\begin{definition}
A \emph{matching function} $F$ maps an element of $[\run(P)]$ to a sub\-se\-quence of $T$. It has to satisfy the following properties for all $i \in [\run(P)]$.
\begin{enumerate}[label=(P\arabic*)]
\item $F(i)$ is a contiguous subsequence of $T$.
\item If the $i$-th run in $P$ is a run up (down), $F(i)$ starts with an element following a valley (peak) or the first element in $T$ and ends with a valley (peak) or the last element in $T$. 
\item $F(1)$ starts with the first and $F(\run(P))$ ends with the last element in $T$.
\item $F(i)$ and $F(i+1)$ have one run in common: $F(i+1)$ starts with the leftmost element in the last run in $F(i)$.
\end{enumerate}
\label{def:matchingfunctions}
\end{definition}

\begin{figure}
\begin{center}
\begin{tikzpicture}[scale=0.7]
	\tikzstyle{up}=[very thick, -, ]
	\tikzstyle{down}=[very thick, dashed, -]
	
	\node  at (-0.5,2.5) {$P$:};	
	\draw[up] (4, 2)--(5, 3); 
	\draw[down] (5, 3)--(6,2);
	\draw[up] (6, 2)--(7, 3); 
	\draw[down] (7, 3)--(8,2);
		
	\node  at (-0.5,0.5) {$T$:};			
	\draw[up] (0, 0)--(1, 1)--(2,0)--(3,1)--(4,0); 
	\draw[down] (3.35,0.9)--(4.25,0)--(5.25,1)--(6.25,0)--(7.25,1)--(8.25,0)--(9.25,1);
	\draw[up] (8.6,0.1)--(9.5,1)--(10.5,0)--(11.5,1)--(12.5,0);
	\draw[down] (11.85,0.9)--(12.75,0)--(13.75,1)--(14.75,0);
	
	\draw[->, thick] (-0.5,2)--(-0.5,1);
	\node  at (-0.75,1.5) {$F$};
	
	\draw[decorate,decoration={brace,amplitude=10}, -] (4,-0.1) -- (0,-0.1);
	\node  at (2,-1) {$=F(1)$};	
	\draw[decorate,decoration={brace,amplitude=10}, -] (9.25,-0.6) -- (3.35,-0.6);
	\node  at (6.125,-1.5) {$=F(2)$};	
	\draw[decorate,decoration={brace,amplitude=10}, -] (12.5,-0.1) -- (8.6,-0.1);
	\node  at (10.25,-1) {$=F(3)$};	
	\draw[decorate,decoration={brace,amplitude=10}, -] (14.75,-0.6) -- (11.85,-0.6);
	\node  at (13.5,-1.5) {$=F(4)$};	

	\end{tikzpicture}
\caption{A sketch of a matching function and its M- and W-shaped subsequences}
\label{fig:ex_Fsketch}
\end{center}
\end{figure}
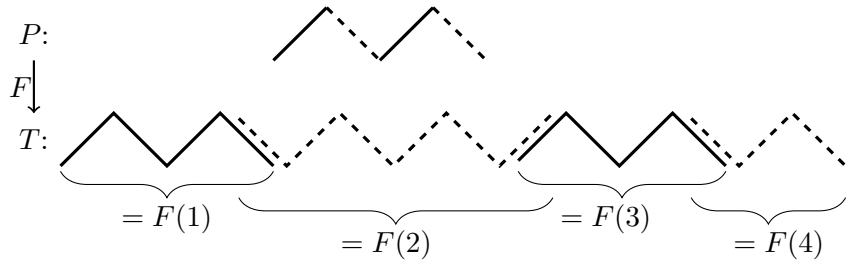

Property (P2) implies that every run up is matched into an M-shaped sequence of runs of the form up--down--up--...--up--down (if the run up is the first or the last run in $P$ the sequence might start or end differently) and every run down is matched into a W-shaped sequence of runs of the form down--up--down--...--down--up (again, if the run down is the first or the last run in $P$, the sequence might start or end differently).
These M- and W-shaped sequences  are sketched in Figure~\ref{fig:ex_Fsketch}.

Property (P4) implies that two adjacent runs in $P$ are mapped to sequences of runs that overlap with exactly one run, as is also sketched in Figure~\ref{fig:ex_Fsketch}.
This overlap is necessary since elements in different runs in $P$ may be matched to elements in the same run in $T$.
More precisely, valleys and peaks in $P$ might be matched to the same run in $T$ as their successors (see the following example).

\begin{example}
\label{ex:matching functions}
In  Figure \ref{fig:ex_F}, $P_{\mathit{ex}}$ (left-hand side) and $T_{\mathit{ex}}$ (right-hand side) are depicted together with a matching function $F$.
A matching compatible with $F$ is given by $4\,6\,2\,9$.
We can see that the elements $6$ and $2$ lie in the same run in $T_{\mathit{ex}}$ even though $3$ (a peak) and $1$ (its successor) lie in different runs in $P_{\mathit{ex}}$.
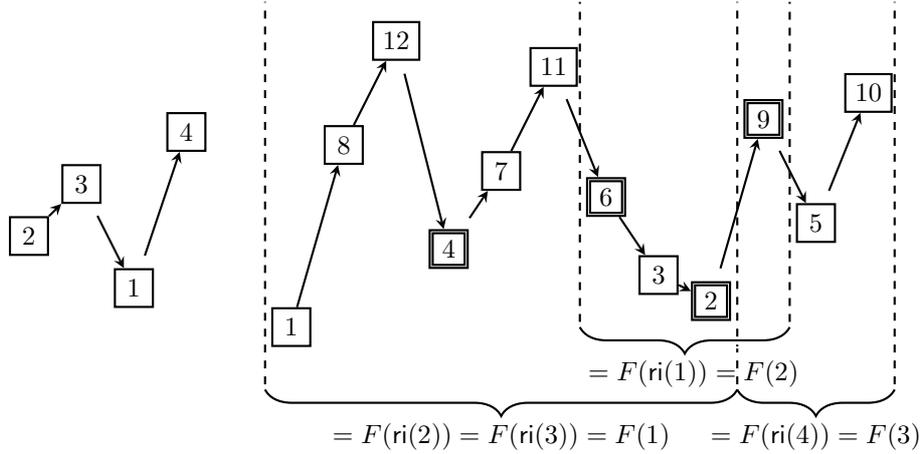
\begin{figure}
\begin{center}
\begin{tikzpicture}[thick, scale=0.69]
	\tikzstyle{every node}=[rectangle,  draw=black, thick, text centered, minimum size= 0.5cm, font=\small] 
	\tikzstyle{m}=[rectangle,  draw=black, thick, text centered, minimum size= 0.42cm, font=\small]
	\tikzstyle{level 2}=[sibling distance=2cm] 
	\pgfsetarrows{-stealth}
	\tikzstyle{a}=[draw=white,text centered, minimum size= 0.7cm, font=\small] 
	\tikzstyle{b}=[rectangle, rounded corners, draw=black, thick, text centered]
	\tikzstyle{1}=[shorten <=5pt]
	\tikzstyle{2}=[dashed, -]
	
	\foreach \x/\y in {1/2, 2/3, 3/1, 4/4}
    \node (p\x) at (\x-5, 0.25+\y) {\y}; 	
  	;
  	\draw (p1) -- (p2);
  	\draw[1] (p2) -- (p3);
  	\draw[1] (p3) -- (p4);

  	\foreach \x/\y in {1/1, 2/8, 3/12, 4/4, 5/7, 6/11, 7/6, 8/3, 9/2, 10/9, 11/5, 12/10}
    \node (t\x) at (\x, 0.5*\y) {\y}; 
    \node[m] at (4,2) {};	
    \node[m] at (7,3) {};
    \node[m] at (9,1) {};
    \node[m] at (10,4.5) {};
  	;
  	
	\foreach \x/\y in {1/2,2/3,5/6,7/8,8/9}
			\draw (t\x) -- (t\y);
	\foreach \x/\y in {3/4,4/5,6/7,9/10,10/11,11/12}
			\draw[1] (t\x) -- (t\y);
			
	\draw[2] (0.5,-0.6) -- (0.5,6.75);
	\draw[2] (6.5,0.6) -- (6.5,6.75);
	\draw[2] (9.5,-0.6) -- (9.5,6.75);
	\draw[2] (10.5,0.6) -- (10.5,6.75);
	\draw[2] (12.5,-0.6) -- (12.5,6.75);
	\node[a] at (5,-1.6) {$=F(\ri(2))=F(\ri(3))=F(1)$};
	\node[a] at (8.5,-0.4) {\ \ $=F(\ri(1))=F(2)$};
	\node[a] at (11,-1.6) {$=F(\ri(4))=F(3)$};
	\draw[decorate,decoration={brace,amplitude=10}, -] (9.5,-0.7) -- (0.5,-0.7);
	\draw[decorate,decoration={brace,amplitude=10}, -] (12.5,-0.7) -- (9.5,-0.7);
	\draw[decorate,decoration={brace,amplitude=10}, -] (10.5,0.5) -- (6.5,0.5);
	\end{tikzpicture}
\caption{$P_{\mathit{ex}}$ and $T_{\mathit{ex}}$ together with a matching function $F$ and the compatible matching witnessed by the subsequence $4\,6\,2\,9$}
\label{fig:ex_F}
\end{center}
\end{figure}
\demo
\end{example}
Note that there are no matching functions if $\run(P)>\run(T)$.
This corresponds to the fact that in such a case no matching from $P$ into $T$ exists either.
The properties (P1)-(P4) guarantee that the number of functions we have to consider is less than ${(\sqrt{2})}^{\run(T)}$, as will be proven in Section~\ref{subsec:runtime}, Lemma~\ref{lem:count functions F}.
This allows us to iterate over all matching functions in FPT time.

Let us formalize what we mean by compatible matchings.

\begin{definition}
A matching $M$ is \emph{compatible} with a matching function $F$ if $M(\kappa)\in F(\ri(\kappa))$ for every $\kappa\in[k]$, i.e., $M$ matches each element contained in the $i$-th run in $P$ to an element in $F(i)$.
\label{def:compatible}
\end{definition}

\begin{lemma}
For every matching $M$ of $P$ into $T$ there exists a matching function $F$ such that $M$ is compatible with $F$.
\label{lem:compatible}
\end{lemma}

The proof of this lemma can be found in Section~\ref{subsec:correctness} on page~\pageref{proof:lem:compatible}.
We continue with the observation that, when searching for a compatible matching by looking for the possible values that $M(1), M(2)$ and so on can take, we do not have to remember \emph{all} the previous choices we made.
Let us have a look at an example first:
\begin{example}
In Figure~\ref{fig:ex_F}, assume that we already have a partial matching: $M(1)=2$ and $M(2)=4$.
We now have to decide where to map $3$.
There are two constraints that have to be satisfied:
First, $M(3)>M(2)$.
Second, $M(3)$ has to be to the right of $M(2)$, since $2 \prec_P 3$.
Since our choices for $M(3)$ are limited to $F(\ri(3))=F(1)$, we do not have to check whether $M(3)$ is left of $M(1)$ but only whether $M(3)>M(2)$.
Later, when deciding where to map $4$, we will only have to verify that $M(4)>M(3)$.

In more generality, we observe that given a matching function and a partial matching $M$ defined on $[\kappa-1]$, deciding where to map $\kappa$ only requires the knowledge of $M(\kappa-1)$ and of $M(\kappa')$, where $\kappa'$ is the previous element in the same run as $\kappa$.
\demo\end{example}

Let us now make this observation more precise:
\begin{lemma}
Let $F$ be a matching function.
A function $M$:$[k] \rightarrow [n]$ is a matching of $P$ into $T$ compatible with $F$ if and only if for every $\kappa\in[k]$:
\begin{enumerate}
\item $M(\kappa)\in F(\ri(\kappa))$,
\item $M(\kappa)>M(\kappa-1)$ and
\item if $\pre(\kappa)$ exists,
then $\pre(\kappa)\prec_P \kappa$ if and only if $M(\pre(\kappa))\prec_T M(\kappa)$, i.e., if $\kappa$ is contained in a run up (down), then $M(\kappa)$ is right (left) of $M(\pre(\kappa))$.
\end{enumerate}
\label{lem:only-predecessors-important}
\end{lemma}

As we will see soon, this lemma is essential for our algorithm. Its proof can be found in Section~\ref{subsec:correctness} on page~\pageref{proof:lem:only-predecessors-important}.

\subsection{Algorithm description}
\label{subsec:algdesc}

Before we start explaining the actual FPT algorithm, let us consider a simple algorithm based on alternating runs.
This simple algorithm (Algorithm~\ref{alg:simpleA}) does not have FPT runtime but has the same basic structure as the FPT algorithm.
In particular, this simple algorithm will already demonstrate the importance of Lemma~\ref{lem:only-predecessors-important}.
\begin{algorithm}
  \SetKw{Or}{or}
  \DontPrintSemicolon
  $X^F_0\leftarrow \{(0, \ldots, 0)\}$\tcp*{The tuple $(0, \ldots, 0)$ has $\run(P)$ elements.}
  \ForEach{matching function $F$}{
  	\For(\tcp*[f]{$\kappa$ is the element to be matched.}){$\kappa \leftarrow 1 \ldots k$} {
  		$X^F_\kappa\leftarrow \emptyset$\;
	  	\ForEach{$\vec{x} \in X^F_{\kappa-1}$} {
			$R\leftarrow \{\nu\in[n]:\nu\in F(\ri(\kappa))\wedge\nu>x_{\ri(\kappa-1)}\wedge (\pre(\kappa)\prec_P \kappa \leftrightarrow x_{\ri(\pre(\kappa))}\prec_T\nu)\}$\tcp*{Conditions according to Lemma~\ref{lem:only-predecessors-important}}
			\ForEach{$\nu \in R$}{
				$X^F_\kappa \leftarrow X^F_\kappa \cup \{(x_1, \ldots, x_{\ri(\kappa)-1}, \nu, x_{\ri(\kappa)+1}, \ldots, x_{\run(P)})\}$\;
			}
		} 
  	}
  	\If{$X^F_k \neq \emptyset$}{
  		\Return{``$P$ can be matched into $T$.''}\;
  	}
  }
  \Return{``$P$ cannot be matched into $T$.''}
\caption{A Simple Alternating Run Algorithm}
\label{alg:simpleA}
\end{algorithm}

From Lemma~\ref{lem:compatible} we know that when checking whether $T$ contains $P$ as a pattern, it is  sufficient to test for all matching functions whether there exists a \emph{compatible} matching.
Let us fix a matching function $F$.
We first find suitable elements to which $1$ can be mapped, then suitable elements for $2$, and so on.
Observe that we can use Lemma~\ref{lem:only-predecessors-important} to verify what suitable elements are.
In addition, Lemma~\ref{lem:only-predecessors-important} tells us that when finding suitable elements for $\kappa \in [k]$, we only require the values of $M(\kappa-1)$ and $M(\pre(\kappa))$.
 This means in particular that we do not have to store all values of a possible partial matching $(M(1),\ldots, M(\kappa))$ but only the values of $M$ for the largest element $\leq \kappa$ in each run in $P$.
For example, when trying to match $P = \perm{2,3,5,7,4,1,6}$ into some text and looking for the possible elements for $\kappa=4$, we only have to consider possibilities for $M(3)$ and $M(\pre(4))=M(1)$.

In this simple algorithm, we want to keep track of all possible partial matchings $(M(1),\ldots, M(\kappa))$ for every $\kappa \in [k]$.
Since such partial matchings can be described by storing a single value per run in $P$, every one of them can be stored as a tuple $\vec{x}$ of length $\run(P)$.
The first element of $\vec{x}$ contains a possible choice for the largest element $\leq \kappa$ in the first run of $P$,
the second element of $\vec{x}$ contains a possible choice for the largest element $\leq \kappa$ in the second run of $P$, etc.
We formalize this notion of ``tuples encoding partial matchings'' as $(\kappa,F)$-matchings:

\begin{definition}
Let $\kappa$ be an integer in $[k]$.
A tuple $\vec{x}=(x_1, x_2, \ldots, x_{\run(P)})$ with $x_i \in [0,n]$ for all $i \in [\run(P)]$ is called a 
$(\kappa, F)$-matching of $P$ into $T$ if the following holds: 
There exists a function $M:[\kappa] \rightarrow [n]$ that is a matching of $P\vert_{[\kappa]}$ into $T$ that is compatible with $F$ and for which it additionally holds that for every ${x}_i\neq 0$, $M(\max\{\kappa'\leq \kappa:\ri(\kappa')=i\})={x}_i$, i.e., $M$ maps the largest element $\leq\kappa$ in the $i$-th run of $P$ to the $i$-th element of $\vec{x}$.
\label{def:(k,F)-matching}
\end{definition}

The following lemma states that $X^F_\kappa$ -- as constructed by  Algorithm~\ref{alg:simpleA} -- indeed contains only tuples that are $(\kappa,F)$-matchings:
\begin{lemma}
Let $X^F_\kappa$ be the set of tuples as constructed by Algorithm~\ref{alg:simpleA}. Then every $\vec{x} \in X^F_\kappa$ is a $(\kappa, F)$-matching.
\label{lem:(k,F)-matchings}
\end{lemma}
The proof can be found in Section~\ref{subsec:correctness} on page~\pageref{proof:lem:(k,F)-matchings}.
As an immediate consequence of this lemma, we know that if $X^F_k\neq\emptyset$ then there exists a matching from $P$ into $T$ that is compatible with $F$.
Observe that $X^F_k$ is always empty if a previous $X^F_\kappa$ was empty.
If for every $F$ the set $X^F_k=\emptyset$, we know from Lemma~\ref{lem:compatible} that $P$ cannot be matched into $T$.

\begin{example}
For our running example $(P_{ex}, T_{ex})$ and $\kappa=1$ the data structure is given as follows: $X^F_1 = \{ (0,6,0), (0,3,0), (0,2,0),(0,9,0)\}$.
Given the choice $M(1)=3$, we obtain $6$ $(2,F)$-matchings, namely: $(8,3,0), (12,3,0),$ $(4,3,0), (7,3,0), (11,3,0)$ and $(6,3,0) $.
In total $X^F_2$ contains 19 elements.
\demo\end{example}

As seen in this small example, the set $R$ and consequently the set $X^F_\kappa$ can get very large.
In particular, it is not possible to bound the size of $X^F_\kappa$ by a function depending only on $\run(T)$ and not on $n$ -- which is necessary for obtaining our FPT result. 
Thus, we have to further refine our algorithm.

We proceed by explaining how this simple algorithm can be improved in order to obtain an FPT algorithm based on alternating runs (Algorithm~\ref{alg:ARA}).
This is the main algorithm described in this paper.
In the following description we fix $F$ to be the current matching function under consideration.
There are two modifications that have to be made in order to obtain FPT runtime.
First, we have to restrict the set $R$ to fewer, \emph{representative} choices.
Second, we have to change the data structure of $X^F_\kappa$ from a set to an array of fixed size.
In the array $X^F_\kappa$, every $(\kappa, F)$-matching has a predetermined position.
Observe that if there are two $(\kappa, F)$-matchings $\vec{x}$, $\vec{y}$ where $\vec{x}$ leads to a matching only if $\vec{y}$ leads to a matching as well, the algorithm only has to remember $\vec{y}$.
The position of a $(\kappa, F)$-matching will thus be assigned in such a way that one of two $(\kappa, F)$-matching sharing the same position is preferable in the above sense.
We will now explain both modifications in detail.

\begin{algorithm}
  \SetKwFunction{Calc}{GetMatching}
  \SetKwFunction{Pos}{Index}
  \SetKwFunction{Minima}{Minima}
  \SetKwFunction{rep}{Rep}
  \SetKwInOut{Input}{input}
  \SetKwInOut{Output}{output}
  \SetKw{Or}{or}
  \DontPrintSemicolon
  $X^F_0\leftarrow[(0, \ldots, 0)]$\tcp*{$(0, \ldots, 0)$ has $\run(P)$ elements.}
  \ForEach{matching function $F$\nllabel{line:cstart}}{
  	\For(\tcp*[f]{$\kappa$ is the element to be matched.}){$\kappa \leftarrow 1 \ldots k$} {
  		$X^F_\kappa\leftarrow[\epsilon, \ldots, \epsilon]$\tcp*{$X^F_\kappa$ is a fixed-size array.}
	  	\ForEach{$\vec{x} \in X^F_{\kappa-1}$ with $\vec{x}\neq \epsilon $} {
	  		$R \leftarrow \rep(\vec{x}, \kappa, F)$\label{line:ARA:6}\;
			\ForEach{$\nu \in R$\label{line:ARA:7}}{
				$i\leftarrow\Pos(x_1, \ldots, x_{\ri(\kappa)-1}, \nu, x_{\ri(\kappa)+1}, \ldots, x_{\run(P)})$\label{line:ARA:8}\;
				$\vec{y}\leftarrow X^F_\kappa(i)$\label{line:ARA:8'}\;
				\If{$\vec{y}=\epsilon$ \Or $y_{\ri(\kappa)}> \nu$\label{line:ARA:9}}{
					$X^F_\kappa(i) \leftarrow (x_1, \ldots, x_{\ri(\kappa)-1}, \nu, x_{\ri(\kappa)+1}, \ldots, x_{\run(P)})$\label{line:ARA:10}\;
				}
			}
		}
  	}
  	\If(\tcp*[f]{Is $X^F_k$ non-empty?}){$X^F_k \neq [\epsilon, \ldots, \epsilon]$}{
  		\Return{``Matching found: \Calc($X^F_1, \ldots, X^F_k$)}''\;
  	}
  }
  \Return{``$P$ cannot be matched into $T$.''}
\caption{The Alternating Run Algorithm}
\label{alg:ARA}
\end{algorithm}

Concerning the first modification, restricting the set $R$, we introduce the procedure $\texttt{Rep}(\vec{x}, \kappa, F)$.
This procedure returns a set of representative elements to which $\kappa$ can be mapped to.
These choices have to be compatible with previously chosen elements ($x_1, x_2, \ldots, x_{\run(P)}$) and the matching function $F$.

\begin{procedure}
  \SetKwFunction{Calc}{Calc}
  \SetKwFunction{Pos}{Pos}
  \SetKwFunction{Val}{Valleys}
  \SetKwInOut{Input}{input}
  \SetKwInOut{Output}{output}
  \SetKw{Or}{or}
  \DontPrintSemicolon
  \Input{a $(\kappa,F)$-matching $\vec{x}=(x_1,x_{2}, \ldots, x_{\run(P)})$, $\kappa\in[k]$, a matching function $F$}
  	\Output{$R$, the set of representative elements for $M(\kappa)$}
  	
  $R \leftarrow F(\ri(\kappa))$\label{line:getRep:1}\;
  $R \leftarrow R \cap [x_{\ri(\kappa-1)}+1,n]$ \label{line:getRep:2a}\;
  $R \leftarrow  \Val(T\vert_R)$ \label{line:getRep:2b}\;
  \eIf{$\kappa$ is in a run up in $P$}{
		\If{ $x_{\ri(\kappa)\neq 0}$ } {
			$R \leftarrow \left\lbrace \nu \in R: x_{\ri(\kappa)} \prec_T  \nu \right\rbrace $\label{line:getRep:3} \;
		}
	    \eIf{$\kappa$ is the largest element in its run}{
			$R \leftarrow \left\lbrace \min R \right\rbrace$\label{line:getRep:4}
		} {
			$R \leftarrow \left\lbrace \nu \in R: \exists \nu' \text{ with } \nu'\in F(\ri(\kappa)) \wedge \nu'>\nu \wedge \nu\prec_T\nu' \right\rbrace$\label{line:getRep:5}
		}
  } {
		\If{ $x_{\ri(\kappa)\neq 0}$ } {
			$R \leftarrow \left\lbrace \nu \in R: \nu \prec_T x_{\ri(\kappa)} \right\rbrace $\label{line:getRep:6} \;
		}
		\eIf{$\kappa$ is the largest element in its run}{
			$R \leftarrow \left\lbrace \min R \right\rbrace$\label{line:getRep:7}
		} {
			$R \leftarrow \left\lbrace \nu \in R: \exists \nu' \text{ with } \nu'\in F(\ri(\kappa)) \wedge \nu'>\nu \wedge \nu'\prec_T\nu \right\rbrace$\label{line:getRep:8}
		}	
  }
  \Return{$R$}			
\caption{Rep($\vec{x}, \kappa,F$)}
\label{alg:getRep}
\end{procedure}

An element $\nu\in[n]$ is contained in $\rep(\vec{x}, \kappa, F)$ if the following conditions are met:

\begin{enumerate}[label=(C\arabic*)]
\item\label{tag:c1} $[$Line~\ref{line:getRep:1}$]$ It has to hold that $\nu\in F(\ri(\kappa))$ (cf.\ Condition 1 in Lemma~\ref{lem:only-predecessors-important}).
\item\label{tag:c2} $[$Line~\ref{line:getRep:2a}$]$ It has to hold that $\nu>x_{\ri(\kappa-1)}$ (cf.\ Condition 2 in Lemma~\ref{lem:only-predecessors-important}).
\item\label{tag:c2'} $[$Line~\ref{line:getRep:2b}$]$
It is always preferable to choose elements that are as small as possible.
To be more precise: If we consider the subsequence of $T$ containing all elements in the set $R$, we  merely need to consider the valleys of this subsequence.
The function $\texttt{Valleys}(T \vert_R)$ returns exactly these valleys.
\item\label{tag:c3} $[$Lines~\ref{line:getRep:3} and \ref{line:getRep:6}$]$  It has to hold that if $\kappa$ is contained in a run up (down), then $\nu$ has to be right (left) of $x_{\ri(\kappa)}$, i.e., the element to which the run predecessor of $\kappa$ is mapped (cf.\ Condition 3 in Lemma~\ref{lem:only-predecessors-important}).
\item\label{tag:c4} $[$Lines~\ref{line:getRep:4} and \ref{line:getRep:7}$]$
If $\kappa$ is the largest element in its run, the optimal choice is the smallest possible element.
\item\label{tag:c5} $[$Lines~\ref{line:getRep:5} and \ref{line:getRep:8}$]$
If $\kappa$ is not the largest element in its run, the choice of $\nu$ must not prevent finding elements for the next elements in its run.
Thus, if $\kappa$ is contained in a run up (down), then there has to be a larger element to its right (left) that is contained in $F(\ri(\kappa))$.
\end{enumerate}

Since this smaller set $R$ is a subset of the set $R$ in the simple algorithm (Algorithm~\ref{alg:simpleA}), we immediately obtain the following corollary of Lemma~\ref{lem:(k,F)-matchings}:
\begin{corollary}
Let $X^F_\kappa$ be the set of tuples as constructed by Algorithm~\ref{alg:ARA}. Then every $\vec{x} \in X^F_\kappa$ is a $(\kappa, F)$-matching.
\label{cor:(k,F)-matchings}
\end{corollary}

\begin{example}
Let us explain how the elements in $\texttt{Rep}((4,2,0),3,F)$ are determined in our running example.
The elements fulfilling Condition~\ref{tag:c1} are: 1, 8, 12, 4, 7, 6, 3 and 2 (listed in the order they appear in $T$).
Among these, the elements larger than $x_{\ri(2)}=x_1=4$ are: 8, 12, 7, 11, 6 (cf.~\ref{tag:c2}). 
If we consider this subsequence, its valleys are: 8, 7, and 6 (these are the elements fulfilling Condition~\ref{tag:c2'}).
The element 3 is contained in a run up in $T$, thus the element it is mapped to has to lie to the right of $x_{\ri(\pre(3)}=x_{\ri(2)}=4$.
The elements also fulfilling \ref{tag:c3} thus are 7 and 6.
Since 3 is the largest element in its run in $P$, we only need to store the smallest possibility which is 6 (cf.~\ref{tag:c4}).
Condition~\ref{tag:c5} does not apply here. 
If there were another, larger element in the same run as 3 in $P$, we would have to choose the element 7, since there are no larger elements in $F(\ri(3))$ to the right of 6.
\demo\end{example}

If any matching of $P$ into $T$ can be found that is compatible with $F$, it is also possible to find a matching that only involves representative elements.
This statement is formalized and proven in Section~\ref{subsec:correctness} (Definition~\ref{def:extending_matchings} and Lemma~\ref{lem:rep.elements-and-extensions}).
For the time being, let us convey the intuition behind this:

\begin{example}
In Figure~\ref{fig:ex_F}, $\{2\mapsto 4,3\mapsto 6,1\mapsto 3,4\mapsto 10\}$ is a matching of $P_{\mathit{ex}}$ into $T_{\mathit{ex}}$ where the elements $3$ and $10$ are not representative: $3 \notin \texttt{Rep}((0,0,0),1,F)$ and $10 \notin \texttt{Rep}((6,3,0),4,F)$.
This can be seen since $3$ is not a valley in $T$ and $10$ is not a valley in the subsequence consisting of elements larger than $6$.
However, this matching can be represented by the matching $\{2\mapsto 4,3\mapsto 6,1\mapsto 2,4\mapsto 9\}$ that only involves representative elements ($3$ is represented by $2$; $10$ by $9$) and that is compatible with the same matching function $F$.
\demo\end{example}

This concludes our description of representative elements, our first modification of the simple alternating run algorithm.
We proceed by explaining the data structure $X^F_\kappa$, which is changed from a set to an array of fixed size.
In this array, every $(\kappa,F)$-matching $\vec{x}$ has a predetermined position which depends on the notion of \emph{vales}.
\begin{definition}
\label{def:vale}
A subsequence of a permutation $\pi$ consisting of a consecutive run down and run up (formed like a V) is called a \emph{vale}.
If $\pi$ starts with a run up, this run is also considered as a vale and
analogously if $\pi$ ends with a run down.
Let $\mathsf{vale}(\pi)$ denote the number of vales in $\pi$.
Finally, we define the \emph{vale index} function $\vi(u)$:
given a matching function $F$ and $u\in F(i)$, let $\vi(u)=j$ if $u$ is contained in the $j$-th vale in $F(i)$.
For notational convenience we set $\vi(0)=1$.
\end{definition}

The main idea is the following: Two $(\kappa,F)$-matchings $\vec{x}$ and $\vec{y}$ in $X^F_\kappa$ with $\vi(x_i)=\vi(y_i)$ for all $i \in [\run(P)]$ are comparable in the sense that one of these is less likely to lead to a matching.
More precisely, the $(\kappa,F)$-matching containing the larger element at the $\ri(\kappa)$-th position (this is also the largest element of the entire tuple) leads to a matching only if the other one leads to a matching as well.
Thus, the former $(\kappa,F)$-matching can be discarded and only the latter $(\kappa,F)$-matching has to be stored.
The following example illustrates this notion of comparability:

\begin{example}
\label{ex:vales}
Consider the two permutations $P$ and $T$ schematically represented in Figure~\ref{fig:min}.
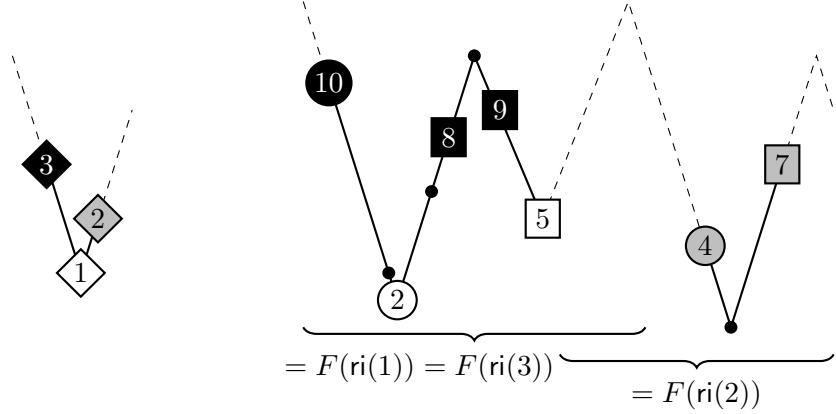
\begin{figure}
\begin{center}
\begin{tikzpicture}[thick, scale=0.45,yscale=0.8]
	\tikzstyle{r2}=[rectangle,  draw=black, fill=white, text centered, minimum size= 0.3cm] 
	\tikzstyle{r1}=[rectangle,  draw=black, fill=black, text centered, text=white, minimum size= 0.3cm]
	\tikzstyle{r3}=[rectangle,  draw=black, fill=black!25, text centered, minimum size= 0.3cm] 
	\tikzstyle{c2}=[circle,  draw=black, fill=white, text centered, inner sep=0.07cm]
	\tikzstyle{c1}=[circle,  draw=black, fill=black, text centered, , text=white,inner sep=0.05cm]
	\tikzstyle{c3}=[circle,  draw=black, fill=black!25, text centered, inner sep=0.07cm]
	\tikzstyle{dot}=[circle,  draw=black, fill=black, text centered, minimum size=0.05cm, inner sep=0.05cm]
	\tikzstyle{d1}=[diamond,  draw=black, fill=black, minimum size=0.1cm,text=white, inner sep=0.05cm]
	\tikzstyle{d2}=[diamond,  draw=black, fill=white, minimum size=0.1cm, inner sep=0.05cm]
	\tikzstyle{d3}=[diamond,  draw=black, fill=black!25, minimum size=0.1cm, inner sep=0.05cm]
	\tikzstyle{2}=[thin, dashed, -]

	\draw (-7,7)--(-6,3)--(-5.5,5);
	\draw[2] (-8,11)--(-7,7);
	\draw[2] (-5.5,5)--(-4.5,9);
	\node[d1] at (-7,7) {3};
	\node[d2] at (-6,3) {1};
	\node[d3] at (-5.5,5) {2};
	
	\draw (1.25,10)--(3.25,2)--(5.5,11)--(7.5,5);
	\draw[2] (0.5,13)--(1.25,10);
	\draw[2] (7.5,5)--(10,13)--(12.25,4);
	\node[c1] at (1.25,10) {10};
	\node[dot] at (3,3) {};
	\node[c2] at (3.25,2) {2};
	\node[dot] at (4.25,6) {};
	\node[r1] at (4.75,8) {8};
	\node[dot] at (5.5,11) {};
	\node[r1] at (6.25,9) {9};
	\node[r2] at (7.5,5) {5};
	\draw[decorate,decoration={brace,amplitude=5}, -] (10.5,1) -- (0.5,1);
		\node at (3.9,-0.5) {$=F(\ri(1))=F(\ri(3))$};
	\draw (12.25,4)--(13,1)--(14.5,7);
	\draw[2] (14.5,7)--(15.5,11)--(16,9);
	\node[c3] at (12.25,4) {4};
	\node[dot] at (13,1) {};
	\node[r3] at (14.5,7) {7};
	\draw[decorate,decoration={brace,amplitude=5}, -] (16,0) -- (8,0);
	\node at (12,-1.5) {$=F(\ri(2))$};

\end{tikzpicture}
\end{center}
\caption{Schematic representation of the permutations occurring in Example~\ref{ex:vales}: 
to the left the pattern $P$, to the right the text $T$.}
\label{fig:min}
\end{figure}
We are searching for representative elements for $\kappa=3$ which lies in a run down in $P$. 
Which elements $\kappa$ may be matched to depends on the choices for its run predecessor $\pre(3)=1$ and for $\kappa-1=2$.
For the element $1$, two representative elements are $2$ (circle) and $5$ (square), the valleys in $F(\ri(1))$ in $T$.
They lead to one representative element for $2$ each: if $2$ has been chosen then $4$ is a representative element (circle) and if $5$ has been chosen then $7$ (square) is one.
At this point, we have the following two $(2,F)$-matchings: $\vec{x}=(\ldots, 0,2,4,0, \ldots)$ and $\vec{y}=(\ldots, 0,5,7,0, \ldots)$.
On the one hand, $\vec{x}$ seems to be preferable since it involves smaller elements than $\vec{y}$ and this leaves more possibilities for the following elements.
On the other hand, $\vec{y}$ seems to be preferable since it involves $5$ in $F(\ri(1))$, which is further to the right than $2$.
This is advantageous since $F(\ri(1))$ corresponds to a run down and this means that larger elements in the same run will have to be chosen to the left.
All together we cannot say which of $\vec{x}$ and $\vec{y}$ is preferable and thus have to store both of them.

When we now turn to the element $3$ in $P$, there are three representative elements: if we have chosen $\vec{x}$ the only possible choice is the element $10$; if we have chosen $\vec{y}$ there are two possible choices namely $8$ and $9$.
We thus obtain three $(3,F)$ matchings: $\vec{x}'=(\ldots, 0,10,4,0, \ldots)$,  $\vec{y}'=(\ldots, 0,8,7,0, \ldots)$ and $\vec{y}''=(\ldots, 0,9,7,0, \ldots)$.
We can now observe that we do not have to keep track of all three possibilities.
Indeed, the two  $(3,F)$-matchings $\vec{x}'$ and $\vec{y}'$ have coinciding vales and $\vec{x}'$ can be discarded in favor of $\vec{y}'$ since $\vec{x}'$ will only lead to a matching of $P$ into $T$ if $\vec{y}'$ does.
This is due to the fact that $x'_{\ri(3)}=10 > 8 = y'_{\ri(3)}$ and can be seen as follows: 

Let $u$ be an element in the same run as $3$ in $P$ that is larger than $3$ (which means that it lies to the left of $3$).
All elements to the left of and larger than $10$ in $F(\ri(u))$ are clearly also to the left of and larger than $8$.
Thus, if there exists an element in $\texttt{Rep}(\vec{x}',u,F)$, then there also exists a smaller element in $\texttt{Rep}(\vec{y}',u,F)$.
This means that from the point of view of the run containing $3$, $\vec{y}'$ is to be preferred over $\vec{x}'$.
Now let $v>3$ be an element in the same run in $P$ as $2$ (which means that it lies to the right of $2$).
Representative elements for $v$ have to both lie to the right of the element chosen for $2$ ($4$ or $7$) and be larger than the element chosen for $3$ ($10$ or $8$).
Since $4$ and $7$ lie in the same vale in $T$ there are no larger elements in between them.
This implies that elements that are to the right of $4$ in $F(\ri(2))$ and larger than $10$ are automatically to the right of $7$ and larger than $8$. 
From the point of view of the run containing $2$, $\vec{y}'$ it also to be preferred over $\vec{x}'$.
The same argument also holds for any other element in $P$ that is larger than $3$.

To put this example in a nutshell: if we have two $(\kappa, F)$-matchings $\vec{x}$ and $\vec{y}$ with coinciding vales and $y_{\ri(\kappa)} \leq x_{\ri(\kappa)}$ we only need to store $\vec{y}$.
For a formal proof of this statement, we refer to Lemma~\ref{lem:index-and-extensions} in Section~\ref{subsec:correctness} on page~\pageref{lem:index-and-extensions}.
\demo
\end{example}

If we store only one $(\kappa,F)$-matching out of those with identical vales, the question arises how many vales there are in $F(i)$, $i\in[\run(P)]$.
The answer is that at most $\lfloor \run(F(i))/2 \rfloor+1$ exist:
all vales but the two outermost consist of two runs and the two outermost may consist of only one run (cf.\ Definition~\ref{def:vale}).
This would yield that we have to store at most $\prod_{i=1}^{\run(P)}\left({\lfloor\frac{\run(F(i)}{2}\rfloor+1}\right)$ many $(\kappa,F)$-matchings.
This number is still too large to show our desired runtime bounds.
However, it suffices to distinguish between $\lfloor\run(F(i))/2\rfloor$ many vales in $F(i)$ with $i\in[\run(P)-1]$.
This is achieved by not distinguishing between the first and the last vale in $F(i)$ for $i < \run(P)$.
We only briefly mention that this is correct due to the Conditions \ref{tag:c4} and \ref{tag:c5}; a formal proof will follow with Lemma~\ref{lem:index-and-extensions} in Section~\ref{subsec:correctness}.
For $i=\run(P)$, the last run in $P$, we still consider all vales occurring in $F(\run(P))$.

Recall that our goal is to assign a position in the array $X^F_\kappa$ to every $(\kappa,F)$-matching $\vec{x}$.
For every one of the $\run(P)$ values of the $(\kappa,F)$-matching there are at most $\lfloor \run(F(i)/2 \rfloor$ vales to be distinguished, except for the last one where we have to take one additional vale into account, i.e., we distinguish between $\lfloor \run(F(\run(P)))/2 \rfloor+1$ vales.
Thus, it is natural to use a mixed radix numeral system with bases $b_1=\lfloor\run(F(1)/2\rfloor, b_2=\lfloor\run(F(2)/2\rfloor, \ldots, b_{\run(P)-1}=\lfloor\run(F(\run(P)-1)/2\rfloor$ and $b_{\run(P)}$ is equal to the number of vales in $F(\run(P))$.
Let \texttt{Index} be the function that assigns a position in the array to each $(\kappa,F)$-matching $\vec{x}=(x_1,\ldots,x_{\run(P)})$:
\begin{align*}
\texttt{Index}\left(x_1,\ldots,x_{\run(P)}\right)=1+\sum_{i=1}^{\run(P)}{(\vi(x_i)-1 \mod b_i) \cdot  \prod_{j=1}^{i-1}{b_j}}.\end{align*}
The $\text{mod}$ operator is required since for $x\in F(i)$, $\vi(x)\in [b_i+1]$ -- as explained above.

\begin{example}
Let us discuss what the \texttt{Index} function looks like for our running example $P_{ex}$ and $T_{ex}$ (cf.\ Figure~\ref{fig:ex_F}).
The subsequence $F(1)$ contains four runs.
Thus, $b_1=2$.
Since both $F(2)$ and $F(3)$ contain two runs, $b_2=b_3=1$.
Consequently, in our running example, $X^F_\kappa$ contains at most two elements for every $\kappa \in [k]$.
For example, $\texttt{Index}(8,3,10)=1$, $\texttt{Index}(6,3,10)=1$ and $\texttt{Index}(11,3,10)=2$.
\demo\end{example}

From the definition of the \texttt{Index}-function, it follows that the length of our array is $\prod_{i=1}^{\run(P)}{b_i}$.
We will show in Lemma~\ref{lem:Xkappa-bound} that $\prod_{i=1}^{\run(P)}{b_i}=\mathcal{O}\left(1.2611^{\run(T)}\right)$.
At this point, we see the huge advantage of this array data structure over the set data structure in the simple algorithm: the set $X^F_\kappa$ has a potential size of $n^{\run{P}}$ -- too large for an FPT algorithm.

This concludes the description of the array data structure.
Let us now -- once again -- return to our running example and see how this would be dealt with by the alternating run algorithm.

\begin{example}
Let us demonstrate how the alternating run algorithm works.
As before, consider $T_{\mathit{ex}}$, $P_{\mathit{ex}}$ and the matching function $F$ as represented in Figure \ref{fig:ex_F}.
We already know from the last example that $X^F_\kappa$ has size $2$, i.e., the \texttt{Index} function has range $\{1,2\}$.
We start with $X^F_0=\{(0,0,0)\}$.
Refer to Table~\ref{tab:example} for an overview.
For the element $1$ in $P$ the only representative element is $2$.
Since $\texttt{Index}(0,2,0)=1$, we store this $(1,F)$-matching at position $1$ in $X^F_1$.
Position $2$ remains empty (symbolized by $\epsilon$).
For the element $2$, we have more representative elements:
$\texttt{Rep}((0,2,0), 2, F)=\{4,8\}$.
Note that $3$ is not a representative element since there is no larger element to its right in $F(\ri(2))=F(1)$ (cf.\ \ref{tag:c5}).
Since $\texttt{Index}(8,2,0)=1$ and $\texttt{Index}(4,2,0)=2$, both $(2,F)$-matchings are stored in $X^F_2$.
For placing the element $3$, observe that $3$ is the largest element in its run in $P$.
Thus, Condition~\ref{tag:c4} applies.
We obtain $\texttt{Rep}((8,2,0), 3, F)=\min\{11,12\}=\{11\}$ as well as $\texttt{Rep}((4,2,0), 3, F)=\min\{7,6\}=\{6\}$.
Thus, we have two $(3,F)$-matchings to store in $X^F_3$: $(11,2,0)$ and $(6,2,0)$
with $\texttt{Index}(11,2,0)=2$ and $\texttt{Index}(6,2,0)=1$.
Finally, we have to place the element $4$.
The $(3,F)$-matching $(11,2,0)$ does not lead to a matching since $\texttt{Rep}((11,2,0), 4, F)=\emptyset$.
However, $\texttt{Rep}((6,2,0), 3, F)=\{9\}$.
Thus, $X^F_4$ contains the $(4,F)$-matching $(6,2,9)$.
This $(4,F)$-matching corresponds to the matching $\{2\mapsto 4,3\mapsto 6,1\mapsto 2,4\mapsto 9\}$.
\begin{table}
\begin{center}
\begin{tabular}{c|c|c|}
& $\texttt{Index}(.,.,.)=1$ & $\texttt{Index}(.,.,.)=2$ \\ 
\hline\rule{0pt}{2.6ex}
$X^F_1$ & $(0,2,0)$ & $\epsilon$ \\ \rule{0pt}{2.6ex}
$X^F_2$ & $(8,2,0)$ & $(4,2,0)$ \\ \rule{0pt}{2.6ex}
$X^F_3$ & $(6,2,0)$ & $(11,2,0)$ \\ \rule{0pt}{2.6ex}
$X^F_4$ & $(6,2,9)$ & $\epsilon$ \\ 
\end{tabular} 
\end{center}
\caption{The arrays $X^F_1,\ldots,X^F_4$ for our running example (cf.\ Figure~\ref{fig:ex_F}).}
\label{tab:example}
\end{table}
\demo\end{example}

Finally, it only remains to explain the \texttt{GetMatching} procedure.
\begin{procedure}
  \SetKw{Or}{or}
  \SetKwInOut{Input}{input}
  \SetKwInOut{Output}{output}
  \DontPrintSemicolon
  \Input{$k$ arrays $X^F_1, X^F_2, \ldots, X^F_k$ generated by Algorithm~\ref{alg:ARA}}
  \Output{$M$, a matching of $P$ into $T$ that is compatible with $F$}
  \For{$\kappa \leftarrow k \ldots 1$}{
  \eIf{$\kappa = k$}{
  		$\vec{x} \leftarrow$ some element in $X^F_k$
		} {
		$\vec{x} \leftarrow$ some element $\vec{y}$ in $X^F_\kappa$ with $x_i=y_i$ for all $i \neq \ri(\kappa)$
		}
  $M(\kappa) \leftarrow x_{\ri(\kappa)}$
  }
  \Return{$M=(M(1), M(2), \ldots, M(k))$}
\caption{GetMatching($X^F_1, \ldots, X^F_k$)}
\label{alg:GetMatch}
\end{procedure}
From Lemma~\ref{lem:(k,F)-matchings} we know that if there is an element in $X^F_k$, a matching from $P$ into $T$ that is compatible with $F$ exists.
However, we have not yet shown how a matching can be constructed from an element in $X^F_k$.
This is what the \texttt{GetMatching} procedure does: it extracts an actual matching $M:[k] \rightarrow [n]$ out of the arrays $X^F_1,\ldots,X^F_k$.
We construct $M$ recursively:
First, we pick some element $\vec{x} \in X^F_k$ and set $M(k)\colonequals x_{\ri(k)}$.
Now, suppose the matching has been determined for $\kappa \in [k]$ and $M(\kappa)= x_{\ri(\kappa)}$ for some $\vec{x} \in X^F_\kappa$.
Then there must exist an element $\vec{y} \in X^F_{\kappa-1}$ that has led to the element $\vec{x}\in X^F_\kappa$, i.e., $\vec{y}$ differs from $\vec{x}$ only at the $\ri(\kappa)$-th element.
We define $M(\kappa-1)\colonequals y_{\ri(\kappa-1)}$.
This defines the function $M: [k]\rightarrow[n]$.
It can easily be seen with the help of Lemma~\ref{lem:only-predecessors-important} that the function $M$ returned by the \texttt{GetMatching} procedure is indeed a matching of $P$ into $T$ that is compatible with $F$.

This concludes our description of the alternating run algorithm. 
We would like to remark that this description omits two minor details necessary for obtaining the polynomial factor $\bigO(n\cdot k)$ of the desired runtime.
The one detail concerns the calculation of the $\texttt{Index}$ function.
The second details concerns how data is stored in the array.
These details are described in the proof of the runtime, Proposition~\ref{prop:fpt runtime}.

\subsection{Correctness}
\label{subsec:correctness}

We start by providing the proof of Lemma~\ref{lem:compatible}, which states that for every matching $M$ there exists a matching function $F$ such that $M$ is compatible with $F$.
\begin{proof}[Lemma~\ref{lem:compatible}]\label{proof:lem:compatible}
Given a matching $M$ from $P$ to $T$, we will construct a matching function $F$ such that $M$ is compatible with $F$.
In order to describe $F$, it is enough to determine the first (=leftmost) element $l_{F(i)}$ of every $F(i)$, where $i\in[\run(P)]$.
In order to specify the last (=rightmost) element $r_{F(i)}$ of $F(i)$ for $i\in[\run(P)]$, we simply need to apply the properties (P3) and (P4): $r_{F(i)}$ is either the last element in $T$ or the leftmost valley (peak) in $F(i+1)$ in case that the $i$-th run is a run up (down).
Clearly, $l_{F(1)}=T(1)$, the first element in $T$ -- cf.~(P3).
When determining $l_{F(i)}$, let $l_{P,i}$ be the first element in the $i$-th run in $P$ and $r_{P,i}$ be the last element in the $i$-th run in $P$.
If the $i$-th run is a run up (down), $l_{F(i)}$ is the right-most element in $T$ lying to the left of or equal to $M(l_{P,i})$ and following a valley (peak).
This construction guarantees that $F$ is a matching function.

In order to prove that $M$ is compatible with $F$, we need to show for all $i\in [\run(P)]$ that $l_{F(i)} \preceq_T M(l_{P,i})$ and $M(r_{P,i}) \preceq_T r_{F(i)}$.
The first statement holds by construction.
For $i=\run(P)$, the second statement clearly also holds by construction.
Let $i \in [\run(P)-1]$.
Let us assume that the $i$-th run is a run up -- the proof for runs down is analogous.
We distinguish between the following cases that are depicted in Figure~\ref{fig:cases_compatible_lemma}:
\begin{figure}
\begin{center}
\begin{center}
\begin{tabular}{|>{\centering\arraybackslash}m{0.29\textwidth}|>{\centering\arraybackslash}m{0.29\textwidth}|>{\centering\arraybackslash}m{0.29\textwidth}|}
\hline
$\diamond$  and $\circ$ in the same run & $\diamond$  and $\circ$ not in the same run; $\circ$ in a run up  & $\diamond$ and $\circ$  not in the same run; $\circ$ in a run down \\ \hline
\vspace{0.3cm}
\begin{tikzpicture}[scale=1]
\tikzstyle{every path}=[thick, -]
\tikzstyle{d}=[thick, dotted]
\tikzstyle{da}=[ thick, dashed]
\tikzstyle{r}=[diamond,  draw=black, fill=white, text centered, text=white, inner sep=0.05cm]
\tikzstyle{c}=[circle,  draw=black, fill=white, text centered, , text=white,inner sep=0.06cm]

\draw (0,0)--(1,1)--(2,0)--(3,1); 
\node[r] at (1.3,0.7) {};
\node[c] at (1.7,0.3) {};
\draw[da] (1.1,-0.5)--(1.1,1.5); 
\draw[d] (2,-1)--(2,1); 
\node at (1.1,-0.75) {$L_{F(i+1)}$};
\node at (2,-1.25) {$R_{F(i)}$};
\end{tikzpicture}  &
\vspace{0.3cm}
\begin{tikzpicture}[scale=1]
\tikzstyle{every path}=[thick, -]
\tikzstyle{d}=[thick, dotted]
\tikzstyle{da}=[ thick, dashed]
\tikzstyle{r}=[diamond,  draw=black, fill=white, text centered, text=white, inner sep=0.05cm]
\tikzstyle{c}=[circle,  draw=black, fill=white, text centered, , text=white,inner sep=0.06cm]

\draw (0,0)--(1,1)--(2,0)--(3,1); 
\node[r] at (1.3,0.7) {};
\node[c] at (2.5,0.5) {};
\draw[da] (1.1,-0.5)--(1.1,1.5); 
\draw[d] (2,-1)--(2,1); 
\node at (1.1,-0.75) {$L_{F(i+1)}$};
\node at (2,-1.25) {$R_{F(i)}$};
\end{tikzpicture} & 
\vspace{0.3cm}
\begin{tikzpicture}[scale=1]
\tikzstyle{every path}=[thick, -]
\tikzstyle{d}=[thick, dotted]
\tikzstyle{da}=[ thick, dashed]
\tikzstyle{r}=[diamond,  draw=black, fill=white, text centered, text=white, inner sep=0.05cm]
\tikzstyle{c}=[circle,  draw=black, fill=white, text centered, , text=white,inner sep=0.06cm]

\draw (0,0)--(1,1)--(2,0)--(3,1); 
\node[r] at (0.6,0.6) {};
\node[c] at (1.7,0.3) {};
\draw[da] (1.1,-0.5)--(1.1,1.5); 
\draw[d] (2,-1)--(2,1); 
\node at (1.1,-0.75) {$L_{F(i+1)}$};
\node at (2,-1.25) {$R_{F(i)}$};
\end{tikzpicture}
\\ \hline
\end{tabular}
\end{center}
\caption{Three cases that have to be distinguished in the proof of Lemma~\ref{lem:compatible} when showing that $M(r_{P,i}) \preceq_T r_{F(i)}$ for all $i \in [\run(P)-1]$ under the assumption that the $i$-th run in $P$ is a run up. The element $M(r_{P,i})$ is represented by a $\diamond$ and the element $M(l_{P,i+1})$ by a $\circ$.}
\label{fig:cases_compatible_lemma}
\end{center}
\end{figure}
\begin{itemize}
\item $M(r_{P,i})$ and $M(l_{P,i+1})$ lie in the same run in $T$. 
Since we have assumed that the $i$-th run in $P$ is a run up, $r_{P,i}$ is a peak in $P$.
Hence, this case is only possible if $M(r_{P,i})$ is in a run down in $T$ and $r_{P,i} > l_{P,i+1}$.
Thus, $l_{F(i+1)}$ is the first element in this run, which implies that $r_{F(i)}$ is the last element of this run and thus $M(r_{P,i}) \preceq_T r_{F(i)}$.
\item $M(r_{P,i})$ and $M(l_{P,i+1})$ do not lie in the same run in $T$ and $M(l_{P,i+1})$ is in a run up in $T$.
In this case, $r_{F(i)}$ is the last element in the run down preceding this run and thus it clearly holds that $M(r_{P,i}) \preceq_T r_{F(i)}$.
\item $M(r_{P,i})$ and $M(l_{P,i+1})$ do not lie in the same run in $T$ and $M(l_{P,i+1})$ is in a run down in $T$.
In this case, $r_{F(i)}$ is the last element in this run and again it clearly holds that $M(r_{P,i}) \preceq_T r_{F(i)}$.\qed
\end{itemize}
\end{proof}

\begin{example}
Constructing $F$ as described in the proof of Lemma~\ref{lem:compatible} for the matching $\perm{4,6,2,9}$ of $P_{ex}$ into $T_{ex}$ yields the matching function represented in Figure~\ref{fig:ex_F}.
\demo\end{example}

Next, we prove Lemma~\ref{lem:only-predecessors-important}.
This lemma states that a function $M$:$[k] \rightarrow [n]$ is a matching of $P$ into $T$ compatible with $F$ if and only if for every $\kappa\in[k]$:
\begin{enumerate}
\item $M(\kappa)\in F(\ri(\kappa))$,
\item $M(\kappa)>M(\kappa-1)$ and
\item if $\pre(\kappa)$ exists,
then $\pre(\kappa)\prec_P \kappa$ if and only if $M(\pre(\kappa))\prec_TM(\kappa)$, i.e., if $\kappa$ is contained in a run up (down), then $M(\kappa)$ is right (left) of $M(\pre(\kappa))$.
\end{enumerate}
\begin{proof}[Lemma~\ref{lem:only-predecessors-important}]\label{proof:lem:only-predecessors-important}
Let $M$:$[k] \rightarrow [n]$ be a matching of $P$ into $T$ that is compatible with $F$.
Recall Definition~\ref{Def.pattern.avoid} which states that $M$ has to be a monotonically increasing function. 
This implies the second condition.
Moreover, the sequence $M(P) = M(P(1)), M(P(2)), \ldots, M(P(k))$ has to be a subsequence of $T$.
This means nothing else than $M(P(1))\prec_T M(P(2)) \prec_T \ldots \prec_T M(P(k))$.
In particular it must hold that $M(u)\prec_T M(v)$, where $u$ and $v$ are two neighbouring elements in the same run in $P$ with $u \prec_P v$.
This implies the third condition.
Finally, the first condition follows directly from the definition of compatibility (Definition~\ref{def:compatible}).

Let $M$:$[k] \rightarrow [n]$ be a function fulfilling the three conditions stated above.
The second condition implies that $M$ is monotonically increasing.
In order to show that $M$ is indeed a matching of $P$ into $T$, we have to show that $M(P) = M(P(1)), M(P(2)), \ldots, M(P(k))$ is a subsequence of $T$.
In other words, we have to show that for all $i \in [k-1]$ it holds that $M(P(i)) \prec_T M(P(i+1))$.
We distinguish three cases:
\begin{itemize}
\item The elements $P(i)$ and $P(i+1)$ lie in the same run in $P$. 
Thus, for the case of a run up (down) we have $P(i)=\pre(P(i+1))$ ($P(i+1)=\pre(P(i))$).
With $\kappa = P(i+1)$ ($\kappa = P(i)$) it follows from the third condition that $M(P(i)) \prec_T M(P(i+1))$ (in both cases). 
\item The elements $P(i)$ and $P(i+1)$ do not lie in the same run in $P$ and $M(P(i))$ and $M(P(i+1))$ do not lie in the same run in $T$.
If $P(i)$ lies in the $j$-th run in $P$, the first condition implies that $M(P(i))$ lies in $F(j)$ and that $M(P(i+1))$ lies in $F(j+1)$ in $T$.
Then property (P4) of matching functions (the leftmost run of $F(j+1)$ is the rightmost run of $F(j)$) implies that $M(P(i))$ lies to the left of $M(P(i+1))$ in $T$.
\item The elements $P(i)$ and $P(i+1)$ do not lie in the same run in $P$ but $M(P(i))$ and $M(P(i+1))$ lie in the same run in $T$.
By the definition of matching functions and since it holds that $M(\kappa)\in F(\ri(\kappa))$ for all $\kappa \in [k]$, this can only be possible if $M(P(i))$ is in the last run of $F(j)$ and $M(P(i+1))$ is in the first run of $F(j+1)$ for some $j \in [\run(P)]$.
Thus, if $P(i)$ lies in a run up (down) in $P$ both $M(P(i))$ and $M(P(i+1))$ are contained in a run down (up) in $T$.
On the other hand, if $P(i)$ is in a run up (down) it must be a peak (valley) and thus it holds that $P(i)>P(i+1)$ ($P(i)<P(i+1)$).
The second condition then ensures that $M(P(i))>M(P(i+1))$ ($M(P(i))<M(P(i+1))$), which implies that $M(P(i))$ lies to the left of $M(P(i+1))$ in $T$.
\end{itemize}
The function $M$ is thus a matching of $P$ into $T$ additionally fulfilling that $M(\kappa)\in F(\ri(\kappa))$ which means that $M$ is a matching compatible with $F$.
\qed\end{proof}

Lemma~\ref{lem:(k,F)-matchings} states that in Algorithm~\ref{alg:simpleA}, $\vec{x} \in X^F_\kappa$ is a $(\kappa, F)$-matching. This can be shown as follows:

\begin{proof}[Lemma~\ref{lem:(k,F)-matchings}]\label{proof:lem:(k,F)-matchings}
We prove this statement by induction over $\kappa$.
For $\kappa=1$ this is easy:
An element $\vec{x} \in X^F_{1}$ looks as follows: $x_i =0$ for all $i \neq \ri(1)$ and $x_{\ri(1)}$ is equal to some $u \in F(\ri(1))$. 
Thus, the function $M: [1] \rightarrow [n]$ with $M(1)=u$ is clearly a $(1, F)$-matching.

Now suppose we have proven the statement of Lemma~\ref{lem:(k,F)-matchings} for $\kappa-1$ and we want to prove it for $\kappa$.
If $\vec{x} \in X^F_\kappa$, then there must exist an element $\vec{y} \in X^F_{\kappa-1}$ and an element $\nu \in [n]$ such that $\vec{x}=(y_1, \ldots, y_{\ri(\kappa)-1}, \nu, y_{\ri(\kappa)+1}, \ldots, y_{\run(P)})$ (see lines 5 to 8 in Algorithm~\ref{alg:simpleA}).
This element $\nu$ may not be any arbitrary element, it must fulfill the following conditions (see Algorithm~\ref{alg:simpleA}, Line 6): $\nu\in F(\ri(\kappa))$, $\nu>x_{\ri(\kappa-1))}$ and $\pre(\kappa)\prec_P \kappa$ if and only if $x_{\ri(\pre(\kappa))}\prec_T\nu$.
Since $\vec{y} \in X^F_{\kappa-1}$ it is a $(\kappa-1, F)$-matching and thus there exists a function $M:[\kappa-1] \rightarrow [n]$ that is a matching of $P\vert_{[\kappa-1]}$ into $T$ that is compatible with $F$ and for which it additionally holds that for every ${y}_i\neq 0$, $M(\max\{\kappa'\leq \kappa-1:\ri(\kappa')=i\})={y}_i$.

We now define a function $\tilde{M}:[\kappa] \rightarrow [n]$ as follows: $\tilde{M}(u)=M(u)$ for all $u\in[\kappa-1]$ and $\tilde{M}(\kappa)= \nu$.
We will see that this function $\tilde{M}$ is a witness for the fact that $\vec{x}$ is a $(\kappa, F)$-matching.
For this purpose we have to check that the three conditions in Lemma~\ref{lem:only-predecessors-important} are fulfilled for every $u \in [\kappa]$.
For $u < \kappa$ these conditions are necessarily fulfilled since we then have $\tilde{M}(u)=M(u)$ and $M$ is a matching of $P\vert_{[\kappa-1]}$ into $T$ that is compatible with $F$.
For $u = \kappa$, i.e., $\tilde{M}(u)=\nu$, these conditions are exactly those stated above that must be fulfilled by the element $\nu \in [n]$.
The last condition in Definition~\ref{def:(k,F)-matching}, namely that for every ${x}_i\neq 0$, $\tilde{M}(\max\{\kappa'\leq \kappa:\ri(\kappa')=i\})={x}_i$, is fulfilled since $M$ is a witness for the fact that $\vec{y}$ is a $(\kappa-1, F)$-matching and since we defined $\tilde{M}(\kappa)$ to be equal to $\nu=x_{\ri(\kappa)}$.
Thus, $\vec{x}$ is a $(\kappa, F)$-matching.
\qed\end{proof}

The next lemma shows that only considering elements returned by the $\texttt{Rep}$ procedure is sound.

\begin{definition}
Let $F$ be a matching function and $\vec{x}=({x}_1, {x}_{2}, \ldots, {x}_{\run(P)})$ be a $(\kappa,F)$-matching for some $\kappa\in[k]$. 
A matching $M$ \emph{$(\kappa, F)$-extends} $\vec{x}$ if $M$ is compatible with $F$ and if for every ${x}_i\neq 0$, $M(\max\{\kappa'\leq \kappa:\ri(\kappa')=i\})={x}_i$, i.e., $M$ maps the largest element $\leq\kappa$ in the $i$-th run of $P$ to the $i$-th element of $\vec{x}$.
\label{def:extending_matchings}
\end{definition}

\begin{definition}
Let $\vec{x} = ({x}_1, \ldots, {x}_{\run(P)})$.
In the following, we write $\vec{x}(\ri(\kappa)\leftarrow \nu)$ instead of $({x}_1, \ldots, {x}_{\ri(\kappa)-1}, \nu, {x}_{\ri(\kappa)+1}, \ldots, {x}_{\run(P)})$.
\label{def:leftarrow-notation}
\end{definition}

\begin{lemma}
Let $\kappa\in[k]$ and $\vec{x}\in X^F_\kappa$.
If there exists a matching $M$ that $(\kappa, F)$-extends $\vec{x}$, then there exist an element 
$\nu \in \texttt{Rep}(\vec{x}, \kappa+1, F)$ and a matching $\tilde{M}$ that $(\kappa+1, F)$-extends $\vec{x}(\ri(\kappa+1)\leftarrow \nu)$.
\label{lem:rep.elements-and-extensions}
\end{lemma}

\begin{proof}
Let us first explicitly show how to pick the element $\nu$.
Then we will prove that it indeed holds that $\nu$ is in $\texttt{Rep}(\vec{x}, \kappa+1, F)$.
We define $\tilde{M}$ as follows: $\tilde{M}(\kappa+1) \colonequals \nu$ and $\tilde{M}(u) \colonequals M(u)$ otherwise.
Finally, we will see that $\tilde{M}$ is a matching  that $(\kappa+1, F)$-extends $\vec{x}(\ri(\kappa+1)\leftarrow \nu)$.

In order to increase legibility, let $i \in [k]$ be the position for which $P(i)=\kappa+1$.
Let us then consider the set ${S}$ consisting of all elements in $T$ that lie to the right of $M(P(i-1))$ and to the left of $M(P(i+1))$, that are contained in $F(\ri(\kappa+1))$ and that are larger than $M(\kappa) = x_{\ri(\kappa)}$.
Thus, 
\[
{S} \colonequals \{ u \in [n]: M(P(i-1)) \prec_T u \prec_T M(P(i+1))\} \cap F(\ri(\kappa+1)) \cap [M(\kappa)+1, n].
\]
This set is never empty:
Especially, $M(\kappa +1)$ is contained in ${S}$ since $M$ is a matching that $(\kappa, F)$-extends $\vec{x}$.
We now define $\nu \colonequals \min({S})$.

We have to check that it indeed holds that $\nu \in \texttt{Rep}(\vec{x}, \kappa+1, F)$. We refer the reader to the definition of $\texttt{Rep}(\vec{x}, \kappa+1, F)$ on page~\pageref{tag:c1}.
\begin{itemize}
\item \ref{tag:c1} is fulfilled by construction of ${S}$.
\item \ref{tag:c2} is fulfilled since $\nu > M(\kappa-1)=x_{\ri(\kappa-1)}$.
\item \ref{tag:c2'} is fulfilled: $\nu$  is a valley in the subsequence of $T$ consisting of elements larger than $M(\kappa)$ by construction of $S$.
\item \ref{tag:c3} If the run predecessor of $\kappa +1$ exists and $\kappa +1$ lies in a run up (down), $\pre(\kappa+1)=P(i-1)$ ($\pre(\kappa+1)=P(i+1)$). Moreover, note that $M(\pre(\kappa+1))=x_{\ri(\kappa+1)}$ since $M$ $(\kappa, F)$-extends $\vec{x}$.
Since ${S} \subseteq \{ u \in [n]: M(P(i-1)) \prec_T u \prec_T M(P(i+1))\}$, it is guaranteed that $\nu$ lies on the correct side of $x_{\ri(\kappa+1)}$.
\item \ref{tag:c4} In case $\kappa +1$ is the largest element in its run in $P$, there is only a single element in $\texttt{Rep}(\vec{x}, \kappa+1, F)$ which is exactly $\nu$.
\item \ref{tag:c5} In case $\kappa +1 $ is not the largest element in its run in $P$ and $\kappa +1$ lies in a run up (down), the element $M(P(i+1))$ ($M(P(i-1))$) is an element larger than $\nu$ that lies to the right (left) of $\nu$ in $F(\ri(\kappa +1))$
since $M$ is compatible with $F$.
\end{itemize}

Now let us show that $\tilde{M}$ as defined above is a matching  that $(\kappa+1, F)$-extends $\vec{x}(\ri(\kappa+1)\leftarrow \nu)$.
First we need to show that the function $\tilde{M}$ is a matching of $P$ into $T$ that is compatible with $F$.
Here Lemma~\ref{lem:only-predecessors-important} comes in handy since it tells us that we only have to check the following three conditions for all $u \in [k]$:
\begin{enumerate}
\item $\tilde{M}(u) \in F(\ri(u))$: For $u=\kappa +1$ this holds by construction of $\nu$ and for $u \neq \kappa+1$ this holds since we then have $\tilde{M}(u)=M(u)$ and $M$ is a matching that is compatible with $F$.
\item  $\tilde{M}(u+1) >  \tilde{M}(u)$ for $u \neq k$: For $u \notin \{\kappa, \kappa +1 \}$ this again holds since $M$ is a matching.

$u = \kappa$: By the construction of ${S}$, $\tilde{M}(\kappa +1) = \nu > M(\kappa)= \tilde{M}(\kappa)$.

$u = \kappa +1$: Again by the construction of ${S}$ we know that $\nu \leq M(\kappa +1)$.
Since $M$ is a matching $M(\kappa +1) < M(\kappa +2) = \tilde{M}(\kappa +2)$ it follows that $\nu = \tilde{M}(\kappa +1) < \tilde{M}(\kappa +2)$.
\item If $\pre(u)$ exists, then $\pre(u)\prec_P u$ if and only if $\tilde{M}(\pre(u))\prec_T \tilde{M}(u)$:
Since $M$ is a matching, we only have to check this condition for $\kappa +1$ and its run predecessor $\pre(\kappa +1)$ as well as for $\kappa +1$ and $\kappa'$, the next largest element in the same run in $P$ (we could call this element the run successor of $\kappa +1$), i.e., $\pre(\kappa')=\kappa+1$.
If $\kappa +1$ lies in a run up (down), we have $\pre(\kappa +1)=P(i-1)$ and $\kappa'=P(i+1)$  ($\pre(\kappa +1)=P(i+1)$ and $\kappa'=P(i-1)$ ).
By construction of ${S}$ we have that $M(P(i-1))= \tilde{M}(P(i-1)) \prec_T \nu=\tilde{M}(\kappa+1) \prec_T \tilde{M}(P(i+1)) = M(P(i+1))$ and thus this condition is also fulfilled.
\end{enumerate}
In order to show that $\tilde{M}$ $(\kappa+1, F)$-extends $\vec{y} \colonequals \vec{x}(\ri(\kappa+1)\leftarrow \nu)$ it remains to show that for every ${y}_i\neq 0$, $\tilde{M}(\max\{\kappa'\leq \kappa:\ri(\kappa')=i\})={y}_i$.
For $i \neq \ri(\kappa +1)$ this follows from the fact that $y_i=x_i$ and that $M$ is a matching that $(\kappa, F)$-extends $\vec{x}$.
For $i = \ri(\kappa +1)$ this hold by definition of $\tilde{M}$: we have $y_i=\nu$ and $\tilde{M}(\max\{\kappa'\leq \kappa+1:\kappa' \text{ is in the same run as } \kappa +1\})=\tilde{M}(\kappa+1)=\nu$.
\qed\end{proof}

It remains to prove that the use of the array data structure and in particular the \texttt{Index} function do not cause that relevant $(\kappa,F)$-matchings are discarded.
This is done by the following two lemmas.

\begin{lemma}
Let $\vec{x},\vec{y}$ be two $(\kappa, F)$-matchings, where $\kappa\in[k]$ and $F$ is a matching function.
If $\texttt{Index}(\vec{x})=\texttt{Index}(\vec{y})$, then for all $i \in [\run(P)]$ it holds that
\begin{itemize}
\item $x_i$ and $y_i$ lie in the same vale in $T$ or 
\item the largest element in the $i$-th run in $P$ is smaller or equal to $\kappa$. 
\end{itemize}
\label{lem:same_index}
\end{lemma}

\begin{proof}
From the definition of the \texttt{Index} function it is clear that $\texttt{Index}(\vec{x})=\texttt{Index}(\vec{y})$ implies that $\vi(x_i) \equiv \vi(y_i) \mod b_i$ for all $i \in [\run(P)]$.
Recall that for $i=\run(P)$, $b_i$ corresponds exactly to the number of vales in $F(i)$ and thus $\vi(x_{\run(P)}) \equiv \vi(y_{\run(P)}) \mod b_i$ is only possible if $\vi(x_{\run(P)}) = \vi(y_{\run(P)})$ which means nothing else than that $x_{\run(P)}$ and $y_{\run(P)}$ lie in the same vale in $T$.

\begin{figure}
\begin{center}
\begin{tabular}{|m{0.7cm}|>{\centering\arraybackslash}m{0.38\textwidth}|>{\centering\arraybackslash}m{0.38\textwidth}|}
\hline
&possible for $i=1$  & possible for $i\in[1,\run(P)-1]$ \\ \hline
\RotText{the $i$-th run in $P$ is a run up} &
\begin{tikzpicture}[scale=0.9]
\tikzstyle{every path}=[very thick, -]
\tikzstyle{d}=[very thick, dotted]
\tikzstyle{da}=[very thick, dashed]
\tikzstyle{every node}=[rectangle,fill=white,text=black, draw=black, thick, text centered, minimum size= 0.05cm]	
\tikzstyle{f}=[rectangle,fill=black,text=black, draw=black, thick, text centered, minimum size= 0.05cm]	

\draw[da] (-0.5,1)--(0,0)--(0.5,1);		
\draw (0.5,1)--(1,0)--(1.5,1); 
\draw[d] (1.7,0)--(2.2,0); 
\draw (2.5,1)--(3,0)--(3.5,1);
\draw[da] (3.5,1)--(4,0);
\end{tikzpicture} 

\vspace*{0.3cm}
$b_1$ = number of vales in $F(1)-1$
 & 
\begin{tikzpicture}[scale=0.9]
\tikzstyle{every path}=[very thick, -]
\tikzstyle{da}=[very thick, dashed]
\tikzstyle{d}=[very thick, dotted]
\tikzstyle{every node}=[rectangle,fill=white,text=black, draw=black, thick, text centered, minimum size= 0.05cm]	
\tikzstyle{f}=[rectangle,fill=black,text=black, draw=black, thick, text centered, minimum size= 0.05cm]	

\draw[da] (1,0)--(1.5,1);			
\draw (1.5,1)--(2,0)--(2.5,1); 
\draw[d] (2.8,0)--(3.2,0); 
\draw (3.5,1)--(4,0)--(4.5,1);
\draw[da] (4.5,1)--(5,0);
\end{tikzpicture}

\vspace*{0.3cm}
$b_i$ = number of vales in $F(i)-1$
\\ \hline
 \RotText{the $i$-th run in $P$ is a run down}  & 
 \begin{tikzpicture}[scale=0.9]
\tikzstyle{every path}=[very thick, -]
\tikzstyle{d}=[very thick, dotted]
\tikzstyle{da}=[very thick, dashed]
\tikzstyle{every node}=[rectangle,fill=white,text=black, draw=black, thick, text centered, minimum size= 0.05cm]	
\tikzstyle{f}=[rectangle,fill=black,text=black, draw=black, thick, text centered, minimum size= 0.05cm]	
		
\draw[da] (1,0)--(1.5,1);
\draw (1.5,1)--(2,0)--(2.5,1); 
\draw[d] (2.8,0)--(3.2,0); 
\draw (3.5,1)--(4,0)--(4.5,1);
\draw[da] (4.5,1)--(5,0)--(5.5,1);
\end{tikzpicture} 

\vspace*{0.3cm}
$b_1$ = number of vales in $F(1)-1$
&
\begin{tikzpicture}[scale=0.9]
\tikzstyle{every path}=[very thick, -]
\tikzstyle{d}=[very thick, dotted]
\tikzstyle{da}=[very thick, dashed]
\tikzstyle{every node}=[rectangle,fill=white,text=black, draw=black, thick, text centered, minimum size= 0.05cm]	
\tikzstyle{f}=[rectangle,fill=black,text=black, draw=black, thick, text centered, minimum size= 0.05cm]	

\draw[da] (0.5,1)--(1,0)--(1.5,1);		
\draw (1.5,1)--(2,0)--(2.5,1); 
\draw[d] (2.8,0)--(3.2,0); 
\draw (3.5,1)--(4,0)--(4.5,1);
\end{tikzpicture} 

\vspace*{0.3cm}
$b_i$ = number of vales in $F(i)$\\
\hline
\end{tabular}
\end{center}
\caption{Possible shapes that $F(i)$ can have in $T$, where $i \neq \run(P)$. 
Runs that are drawn with dashed lines indicate that elements $x$ lying in these runs fulfil $\vi(x) \equiv 1 \mod b_i$.}
\label{fig:fourcases}
\end{figure}
For the case that $i \neq \run(P)$, this is not always that simple.
Consider the four possible shapes that $F(i)$ can have, as depicted in Figure~\ref{fig:fourcases}.
Let us first take a look at the case that the $i$-th run in $P$ is a run up.
Here, $\vi(x_i) \equiv \vi(y_i) \mod b_i$ is possible if $x_i$ and $y_i$ lie in the same vale in $T$ or if $x_i$ lies in the first vale in $F(i)$ and $y_i$ lies in the last run in $F(i)$ (or vice-versa).
Now recall the definition of the \texttt{Rep} procedure: an element in the last run (which is always a run down) may only be chosen for the largest element in its run in $P$ (Condition \ref{tag:c5}).
This means that the largest element in the $i$-th run in $P$ must be smaller or equal to $\kappa$.
Now let us consider the case that the $i$-th run in $P$ is a run down.
Here, if $x_i$ and $y_i$ do not lie in the same vale in $T$, $\vi(x_i) \equiv \vi(y_i) \mod b_i$ is only possible for $i=1$ and if $T$ starts with a run up: $x_i$ has to then lie in this first run of $T$ and $y_i$ in the last vale of $F(1)$ (or vice-versa).
Again, because of Condition \ref{tag:c5}, this is only possible for the largest element in its run in $P$.
Thus, we can again conclude that the largest element in the $i$-th run in $P$ must be smaller or equal to $\kappa$.
\end{proof}

\begin{lemma}
Let $\vec{x},\vec{y}$ be two $(\kappa, F)$-matchings, where $\kappa\in[k]$ and $F$ is a matching function.
In addition to that, let $\nu_x\in \texttt{Rep}(\vec{x}, \kappa+1, F)$ and $\nu_y\in \texttt{Rep}(\vec{y}, \kappa+1, F)$.
If 
\[\texttt{Index}(\vec{x}(\ri(\kappa+1)\leftarrow \nu_x))=\texttt{Index}(\vec{y}(\ri(\kappa+1)\leftarrow \nu_y))\]
and $\nu_y \leq \nu_x$ the following holds:
if there exists a matching that $(\kappa+1,F)$-extends $\vec{x}(\ri(\kappa+1)\leftarrow \nu_x)$, then there exists a matching that $(\kappa+1,F)$-extends $\vec{y}(\ri(\kappa+1)\leftarrow \nu_y)$.
Thus, the alternating run algorithm only has to keep track of the $(\kappa+1, F)$-matching $\vec{y}(\ri(\kappa+1)\leftarrow \nu_y)$.
\label{lem:index-and-extensions}
\end{lemma}

\begin{proof}
Let $M_x$ be a matching of $P$ into $T$ that $(\kappa+1,F)$-extends $\vec{x}(\ri(\kappa+1)\leftarrow \nu_x)$. 
We shall construct a function $M_y:[k]\rightarrow [n]$ and show that it is a matching that $(\kappa+1,F)$-extends $\vec{y}(\ri(\kappa+1)\leftarrow \nu_y)$.

Since $\vec{y}$ is a  $(\kappa, F)$-matching (Recall Definition~\ref{def:(k,F)-matching}) there exists a partial matching $M:[\kappa] \rightarrow [n]$ of $P\vert_{[\kappa]}$ into $T$ for which it additionally holds that for every ${y}_i\neq 0$, $M(\max\{\kappa'\leq \kappa:\ri(\kappa')=i\})={y}_i$.
We define the function $M_y$ as follows:
\[
M_y(u) = \begin{cases} M(u), &\text{ for } u \in [\kappa] \\
\nu_y, &\text{ for } u = \kappa +1 \\
M_x(u), &\text{ for } u \in [\kappa+2, k] \\
\end{cases}
\]

We now need to show that $M_y$ is indeed a matching that $(\kappa+1,F)$-extends $\vec{y}(\ri(\kappa+1)\leftarrow \nu_y)$.
As in the proof of Lemma~\ref{lem:rep.elements-and-extensions}, we shall use Lemma~\ref{lem:only-predecessors-important} to show that $M_y$ is a matching that is compatible with $F$.
We have to check the following three conditions for all $u \in [k]$:
\begin{enumerate}
\item $M_y(u) \in F(\ri(u))$: For $u=\kappa +1$ this holds since $\nu_y \in \texttt{Rep}(\vec{y}, \kappa+1, F)$ (Condition~\ref{tag:c1}) and for $u \neq \kappa+1$ this holds since $M_x$ and $M$ are matchings that are compatible with $F$.
\item  $M_y(u+1) >  M_y(u)$ for $u \neq k$: For $u \notin \{\kappa, \kappa +1 \}$ this again holds since $M_x$ and $M$ are matchings.
\begin{enumerate}
\item $M_y(\kappa +1) > M_y(\kappa)$ or equivalently $\nu_y > M(\kappa)=y_{\ri(\kappa)}$: This holds since $\nu_y \in \texttt{Rep}(\vec{y}, \kappa+1, F)$ (Condition~\ref{tag:c2}).
\item $M_y(\kappa +2) > M_y(\kappa+1)$ or equivalently $M_x(\kappa+2) > \nu_y$: Since $M_x$ is a matching that $(\kappa+1,F)$-extends $\vec{x}(\ri(\kappa+1)\leftarrow \nu_x)$ it has to hold that $M_x(\kappa +2) > M_x(\kappa+1)= \nu_x$.
Since we have $\nu_y \leq \nu_x$, this condition is fulfilled.
\end{enumerate}
\item If $\pre(u)$ exists, then $\pre(u)\prec_P u$ if and only if $M_y(\pre(u))\prec_T M_y(u)$:
Since $M_x$ and $M$ are matchings, this condition is fulfilled for all $u \in [k]$ such that both $u< \kappa+1$ and $\pre(u) < \kappa+1$ or such that both $u > \kappa+1$ and $\pre(u) > \kappa+1$.
Thus, we only have to check this condition for $u=\kappa+1$ and for all $\kappa'\in[\kappa+2,k]$ that satisfy $\pre(\kappa')\leq\kappa+1$.
Let $K$ be the set of all such $\kappa'$.
Observe that such a $\kappa'$ is the smallest element in the $\ri(\kappa')$-th run in $P$ that is strictly larger than $\kappa+1$.
This means that $\pre(\kappa')$, if it exists, is the largest element in the $\ri(\kappa')$-th run in $P$ that is smaller or equal to $\kappa+1$.
We only consider the case that $u$ is contained in a run up --  
the proof for the case that $u$ lies in a run down works analogously.
We have to check the condition for the following three situations:
\begin{enumerate}
\item $u=\kappa+1$: If $\pre(\kappa+1)$ exists it has to hold that $M_y(\pre(\kappa+1)) = y_{\ri(\kappa+1)} \prec_T \nu_y$.
This condition is fulfilled since $\nu_y \in \texttt{Rep}(\vec{y}, \kappa+1, F)$ (Condition~\ref{tag:c3}).
\item $u=\kappa'\in K$ such that $\pre(\kappa')=\kappa+1$: 
If this element $\kappa'$ exists we have to show that $\nu_y \prec_T M_y(\kappa')=M_x(\kappa')$. 
Since $\kappa+1$ is not the largest element in its run in $P$, we know from Lemma~\ref{lem:same_index} that $\nu_x$ and $\nu_y$ lie in the same vale in $T$.
Moreover we know that $\nu_x \geq \nu_y$ -- but what does this imply for the right-left order of $\nu_x$ and $\nu_y$ within this vale? Two cases may occur: $\nu_x$ may lie in the run up or in the run down of this vale.
If $\nu_x$ lies in the run up, then it has to hold that $\nu_y \prec_T \nu_x$.
Since $M_x$ is a matching, it has to hold that $\nu_x = M_x(\kappa+1)\prec_T M_x(\kappa')$ and thus $\nu_y \prec_T M_x(\kappa')$.
If $\nu_x$ lies in the run down, $\nu_x \prec_T \nu_y$ and all elements between $\nu_x$ and $\nu_y$ in $T$ are smaller than $v_x$.
This implies that $M_x(\kappa')$ which is larger than $\nu_x$ and lies to the right of $\nu_x$ also has to lie to the right of $\nu_y$ in $T$.
\item $u=\kappa'\in K$ with $\pre(\kappa')<\kappa+1$: 
We need to show that $y_{\ri(\pre(\kappa'))} =y_{\ri(\kappa')} =M(\pre(\kappa')) \prec_T M_x(\kappa')$.
Since $M_x$ is a matching that $(\kappa+1,F)$-extends $\vec{x}(\ri(\kappa+1)\leftarrow \nu_x)$, we know that $M_x(\pre(\kappa'))=x_{\ri(\kappa')}$ and that $x_{\ri(\kappa')} \prec_T M_x(\kappa')$.
Moreover, since $\texttt{Index}(\vec{x}(\ri(\kappa+1)\leftarrow \nu_x))=\texttt{Index}(\vec{y}(\ri(\kappa+1)\leftarrow \nu_y))$ and $\pre(\kappa')$ is not the largest element in its run in $P$ we know from Lemma~\ref{lem:same_index} that $x_{\ri(\kappa')}$ and $y_{\ri(\kappa')}$ lie in the same vale in $T$.
However, nothing is known about the relative positions of these two elements within this vale and we have to distinguish two cases.
If $y_{\ri(\kappa')} \prec_T x_{\ri(\kappa')}$ the statement follows easily since $y_{\ri(\kappa')} \prec_T x_{\ri(\kappa')} \prec_T M_x(\kappa')$.
If $x_{\ri(\kappa')} \prec_T y_{\ri(\kappa')}$ we have to collect a few more arguments in order to prove that the condition holds.
By transitivity and the condition checked in Point 2.\ of this proof we know that $y_{\ri(\kappa')} < \nu_y=M_y(\kappa+1) < M_x(\kappa')$.
Now note that the elements that lie in $T$ between $x_{\ri(\kappa')}$ and $y_{\ri(\kappa')}$ are all smaller than $\max(x_{\ri(\kappa')},y_{\ri(\kappa')})$ (since both are contained in the same vale).
Thus, the element $M_x(\kappa')$ -- that is to the right of $x_{\ri(\kappa')}$ and larger than $y_{\ri(\kappa')}$ -- has to lie to the right of $y_{\ri(\kappa')}$. 
This is what we wanted to prove.
\end{enumerate}
\end{enumerate}

Let $\vec{y}'=\vec{y}(\ri(\kappa+1)\leftarrow \nu_y)$.
It remains to show that for every $i\in[\run(P)]$ with $y'_i\neq 0$, $M_y(\max\{\kappa'\leq \kappa+1:\ri(\kappa')=i\})=y'_i$.
This follows directly from the definition of $M_y(\kappa +1)$ and the fact that $M$ is a witness for $\vec{y}$ being a $(\kappa,F)$-matching.
\qed\end{proof}

Finally, we have gathered all necessary information to prove the correctness of the alternating run algorithm.

\begin{proposition}
$P$ can be matched into $T$ if and only if $X^F_k$ is non-empty for some matching function $F$.
\end{proposition}
\begin{proof}
($\Rightarrow$) If there is a matching of $P$ into $T$, then there is at least one matching function $F$ for which $X^F_k$ is nonempty:\\
Since there exists a matching $M$, we know from Lemma~\ref{lem:compatible} that there exists some matching function $F$ such that $M$ is compatible with $F$.
Let us fix this $F$.
We prove by induction over $\kappa \in [k]$ that there is an $\vec{x}\in X^F_\kappa$ and a matching $M_\kappa$ that $(\kappa,F)$-extends $\vec{x}$.
For $\kappa=1$ this is easy.
Let $\nu$ be the valley in $T$ that lies in the same vale as $M(1)$.
It is clear that $\nu\in \texttt{Rep}((0,\ldots,0),1,F)$.
Consequently, the tuple $\vec{x}$ with $x_i=0$ for $i\neq \ri(1)$ and $x_{\ri(1)}=\nu$ is contained in $X_1^F$.
Observe that $M_1$ being defined by $M_1(u)=M(u)$ for $u\neq 1$ and $M_1(1)=\nu$ is a matching that $(1,F)$-extends $\vec{x}$.

Now, let $\kappa\in[k]$ and assume that $\vec{x}\in X^F_\kappa$ and $M_\kappa$ $\kappa$-extends $\vec{x}$.
We show that there exist an $\vec{x}'\in X^F_{\kappa+1}$ and a $M_{\kappa+1}$ that $(\kappa+1)$-extends $\vec{x}'$.
By Lemma~\ref{lem:rep.elements-and-extensions}, there exists a $\nu\in \texttt{Rep}(\vec{x}, \kappa+1, F)$ and a matching $M_{\kappa+1}$ that $(\kappa+1)$-extends $\vec{x}(\ri(\kappa+1)\leftarrow\nu)$.
At this point, we cannot be sure that $\vec{x}(\ri(\kappa+1)\leftarrow\nu)\in X^F_{\kappa+1}$ since $X^F_{\kappa+1}$ may contain another $(\kappa,F)$-matching $\vec{y}$ with $\texttt{Index}(\vec{x})=\texttt{Index}(\vec{y})$.
However, this is only possible if $y_{\ri(\kappa +1)} \leq x_{\ri(\kappa +1)}$ (see Line~\ref{line:ARA:9} in Algorithm~\ref{alg:ARA}). 
By Lemma~\ref{lem:index-and-extensions} we know that, in this case, there exists a matching that $(\kappa+1)$-extends $\vec{y}$.
So, no matter whether $\vec{x}(\ri(\kappa+1)\leftarrow\nu)\in X^F_{\kappa+1}$ or not, we can conclude that there is an $\vec{x}'\in X^F_{\kappa+1}$ and a matching function $M_{\kappa+1}$ that $(\kappa+1)$-extends $\vec{x}'$.
By induction, we have shown that $X^F_k\neq \emptyset$.

\noindent ($\Leftarrow$) If there is a matching function $F$ such that the corresponding $X^F_k$ is non-empty, then a matching of $P$ into $T$ can be found: This is an immediate consequence of Corollary~\ref{cor:(k,F)-matchings}.
\qed\end{proof}

Finally, let us remark that the function $M$ returned by $\texttt{GetMatching}(X^F_1,\ldots,X^F_k)$ is indeed a matching, as can easily be seen with the help of Lemma~\ref{lem:only-predecessors-important}:
The first condition in the lemma is satisfied because of Condition~\ref{tag:c1} for representative elements.
The second condition holds because of Condition \ref{tag:c2}.
The third condition corresponds to Condition \ref{tag:c3}.
Note that\ref{tag:c2'},  \ref{tag:c4} and \ref{tag:c5} are only required for improving the runtime.

\subsection{Runtime} 
\label{subsec:runtime}

We are now going to prove the promised FPT runtime bounds.
First, we bound the number of matching functions.
\begin{lemma}
There are less than ${(\sqrt{2})}^{\run(T)}$ functions from $[\run{P}]$ to subsequences of $T$ that satisfy \textit{(P1)} to \textit{(P4)}.
\label{lem:count functions F}
\end{lemma}
\begin{proof}
A matching function $F$ can be uniquely characterized by fixing the position of the first run up in every $F(i)$ for $i \in [\run(P)]$.
This is because the last run of $F(i)$ is the first run of $F(i+1)$ for all $i \in [\run(P)-1]$. 
Moreover the first run up in $F(1)$ is always the first run up in $T$.  
Thus, the number of matching functions is equal to the number of possibilities of picking  $\run(P)-1$ runs (for the first run in $P$ no choice has to be made) among the at most $\lceil\run(T)/2\rceil$ runs up in $T$. 
Hence, we obtain
\begin{equation*}
\binom{\lceil\run(T)/2\rceil}{\run(P)-1} 
\leq 2^{\lceil\run(T)/2\rceil -1} < {(\sqrt{2})}^{\run(T)}.
\end{equation*}
The first inequality holds since $\binom{n}{k} < 2^{n-1}$ for all $n, k \in \mathbb{N}$ as can easily be proven by induction over $n$. 
\qed\end{proof}

Now we bound the size of $X^F_\kappa$, which is the main step to achieve the $1.79^{\run(T)}$ runtime bound.

\begin{lemma}
\label{lem:Xkappa-bound}
For any given matching function $F$ and every $\kappa\in[k]$ 
\begin{equation*}
\card{X^F_\kappa} \leq 2\cdot \prod_{i=1}^{run(P)}{\run(F(i))\over 2} \leq 1.6 \cdot 1.261071^{\run(T)}.
\end{equation*}
\end{lemma}

\begin{proof}
Recall that each $(\kappa,F)$-matching in $X^F_\kappa$ has a position as determined by the function \texttt{Index}, defined by
\begin{align*}
\texttt{Index}(x_1,\ldots,x_{\run(P)})=1+\sum_{i=1}^{\run(P)}{(\vi(x_i) \mod b_i) \cdot  \prod_{j=1}^{i-1}{b_j}}.
\end{align*}
For $i\in [\run(P)-1]$, $b_i= \lfloor\run(F(i)/2\rfloor$, and $b_{\run(P)}\leq\lfloor\run(F(\run(P))/2\rfloor+1$ since $b_{\run(P)}$ is equal to the number of vales in $F(\run(P))$
\footnote{The reason why we do not set $b_{\run(P)}=\lfloor\run(F(\run(P))/2\rfloor$ is a rather technical one: $F(\run(P))$ may end with a run up if the last run in $P$ is a run up and may end with a run down if the last run in $P$ is a run down.
This would lead to unwanted collisions concerning the \texttt{Index} function and consequently would prohibit the proof of Lemma~\ref{lem:same_index}.}.
The range of \texttt{Index} is $\left\{1,\ldots,\prod_{i=1}^{\run(P)}b_i\right\}$.
Since the function \texttt{Index} determines the positions in the array $X^F_\kappa$, we obtain \[\card{X^F_\kappa}= \prod_{i=1}^{\run(P)}b_i \leq \prod_{i=1}^{\run(P)-1}\left\lfloor{\run(F(i))\over 2}\right\rfloor \cdot \left(\left\lfloor{\run(F(\run(P)))\over 2}\right\rfloor+1\right)\]
and consequently
\begin{align}
\card{X^F_\kappa}\leq 2\cdot \prod_{i=1}^{\run(P)}{\run(F(i))\over 2}.
\label{eq:Xkappa-runs}
\end{align}
We want to bound $X^F_\kappa$ and thus want to know when the product in Equation~\eqref{eq:Xkappa-runs} is maximal.
The maximum of this product has to be determined under the condition that
\begin{equation}
\sum_{i=1}^{\run(P)}{\run(F(i))}=\run(T)+\run(P)-1,
\label{eqn:constr}
\end{equation}
since two subsequent $F(i)$'s have one run in common (cf.\ Definition~\ref{def:matchingfunctions}).
The inequality of geometric and arithmetic means implies that the product in Equation~\eqref{eq:Xkappa-runs} is maximal if all $\run(F(i))$ are equal, i.e., for every $i \in \run(P)$, 
$\run(F(i))=\frac{\run(T)+\run(P)-1}{\run(P)}$. 
Therefore, $X^F_\kappa$ has at most $2\cdot\left(\frac{\run(T)+\run(P)-1}{2\cdot \run(P)}\right)^{\run(P)}$ elements.
To shorten the proof, we write in the following $p$ for $\run(P)$ and $t$ for $\run(T)$.
Thus, we want to determine the maximum of the function 
\[g(p)=\left( \frac{t+p-1}{2p}\right)^p\]
(we omit the factor $2$ for the calculation).
\begin{align*}
&g'(p)=\frac{1}{p}\Bigg( 2^{-p}\Big( \frac{p+t-1}{p}\Big)^{p-1} \cdot\\
&\cdot\Big( (p+t-1)\log\Big( \frac{p+t-1}{p}\Big) -p\log(2)-t(1+\log(2)) + 1+ \log(2) \Big)\Bigg) \stackrel{\mathrm{!}}=0
\end{align*}
\[\Longrightarrow (p+t-1)\left( \log \left( \frac{p+t-1}{p}\right) - \log(2)\right)-t+1=0 \]
\[\Longrightarrow  \log \left( \frac{p+t-1}{2p} \right)=\frac{t-1}{p+t-1}. \]
The solutions are:
\begin{eqnarray*}
p_1(t) &=& (-1+t)/(-1+2 e^{1+W_{0}(-1/(2 e))}) \\
p_2(t) &=& (-1+t)/(-1+2 e^{1+W_{-1}(-1/(2 e))}),
\end{eqnarray*}
where $W_{0}$ is the principal branch of the Lambert function (defined by $x=W(x)\cdot e^{W(x)}$) and $W_{-1}$ its lower branch.
It holds that
\begin{eqnarray*}
(-1+t)/3.311071 &\leq & p_1(t)  \leq  (-1+t)/3.311070 \\
(-1+t)/-0.62663 &\leq & p_2(t)  \leq(-1+t)/-0.62664,
\end{eqnarray*}
The second solution $p_2(t)$ is negative and therefore of no interest to us.
The first solution $p_1(t)$ is a local maximum as can be checked easily and yields
\begin{align*}
 g(p_1) & \leq \left( \frac{t+(-1+t)/3.311070-1}{2(-1+t)/3.311071}\right)^{(-1+t)/3.311070} 
 \leq 0.80 \cdot \left( 1.261071\right) ^t.
\end{align*}
It therefore holds that $|X^F_\kappa| \leq 1.6\cdot 1.261071^{\run(T)}$.
\qed\end{proof}

\begin{proposition}
\label{prop:fpt runtime}
The runtime of the alternating run algorithm is $\bigO(1.784^{\run (T)}\cdot n\cdot k)$.
\end{proposition}

\begin{proof}
The main structure of the algorithm is the following:
for every matching function $F$ and for every $\kappa\in[k]$ the array $X^F_\kappa$ is computed.
There are ${(\sqrt{2})}^{\run(T)}$ matching functions (Lemma~\ref{lem:count functions F}).
The maximal number of elements in $X^F_\kappa$ is $1.6\cdot 1.2611^{\run(T)}$ (Lemma~\ref{lem:Xkappa-bound}).
Given a matching function and an element $\kappa\in [k]$, the algorithm has to execute Lines~\ref{line:ARA:6} to \ref{line:ARA:10} for every $\vec{x}\in X^F_{\kappa-1}$.
Once we have shown that the runtime of these lines is $\mathcal{O}(n)$, we obtain a total runtime of $\bigO\left({(\sqrt{2})}^{\run(T)}\cdot 1.2611^{\run (T)}\cdot k\cdot n\right)=\bigO(1.784^{\run (T)}\cdot k\cdot n)$.

So it remains to show that the runtime of the Lines~\ref{line:ARA:6} to \ref{line:ARA:10} is $\mathcal{O}(n)$.
First, observe that determining the set $R$ with the help of the \texttt{Rep} procedure requires $\mathcal{O}(n)$ time.
Second, for every element in $R$ the Lines~\ref{line:ARA:8}, \ref{line:ARA:9} and \ref{line:ARA:10} are executed.
Since $R$ only contains valleys (of some subsequence of $T$), its size is less than $\run(T)$.
Since we assume constant costs for arithmetic operations, computing \texttt{Index} requires $\mathcal{O}(\run(P))$ time.
However, note that it is not necessary to repeat all calculations for \texttt{Index} for every element $\nu$ in $R$.
Indeed, for a fixed $\vec{x} \in X^F_\kappa$, the elements for which \texttt{Index} is computed at Line~\ref{line:ARA:8} only differ at the $\ri(\kappa)$-th position.
Assume that we have already computed \texttt{Index}$(\vec{x})$ for some $\vec{x}$.
Computing \texttt{Index}$(\vec{y})$ for a $\vec{y}$ that is identical to $\vec{x}$ except at the $\ri(\kappa)$-th position can be done as follows:
\begin{align*}
\texttt{Index}(\vec{y})=\texttt{Index}(\vec{x})+\left(\vi(y_{\ri(\kappa)})-\vi(x_{\ri(\kappa)}) \mod b_{\ri(\kappa)}\right) \cdot  \prod_{j=1}^{{\ri(\kappa)}-1}{b_j}.
\end{align*}
Consequently, Line~\ref{line:ARA:8} requires (amortized) constant time.

Checking the condition in Line~\ref{line:ARA:9} requires only constant time.
However, Line~\ref{line:ARA:10} requires $\mathcal{O}(\run(P))$ time to write the $(\kappa,F)$-matching to its position in $X^F_\kappa$.
This is too much time to obtain the desired runtime bound --
we can only afford amortized $\bigO(n)$ time per $\vec{x}\in X^F_{\kappa-1}$.
This can be achieved by executing Line~\ref{line:ARA:10} at most once per $\vec{x}\in X^F_{\kappa-1}$.
Let $\nu \in R$ be the first element for which the condition at Line~\ref{line:ARA:9} is fulfilled. 
For this element Line~\ref{line:ARA:10} is executed and a pointer $p'$ to the position $\texttt{Index}(\vec{x}(\ri(\kappa) \leftarrow \nu))$ is created.
(Recall the $\vec{x}(\ri(\kappa) \leftarrow \nu)$ notation from Definition~\ref{def:leftarrow-notation}.)
If the condition at Line~\ref{line:ARA:9} is fulfilled for the same $\vec{x}$ and some other $\nu' \in R$, we do not execute Line~\ref{line:ARA:10}.
Instead we only store the pointer $p'$ and the element $\nu'$.
This is sufficient information since two $(\kappa,F)$-matchings in Line~\ref{line:ARA:10} that originate from the same $\vec{x}$ are identical except for the $\ri(\kappa)$-th element.
It might be that Line~\ref{line:ARA:10} is executed for some other element $\vec{y} \in X^F_{\kappa-1}$ and $\nu_y \in \texttt{Rep}(\vec{y}, \kappa, F)$ at a later point.
It is then possible that a $(\kappa,F)$-matching $\vec{x}(\ri(\kappa) \leftarrow \nu)$ is overwritten that has other $(\kappa,F)$-matchings $\vec{x}(\ri(\kappa) \leftarrow \nu')$ pointing to it.
However, this can only happen in the following situation: $\vec{x}(\ri(\kappa) \leftarrow \nu')$ is $(\kappa,F)$-extendable only if $\vec{y}(\ri(\kappa) \leftarrow \nu')$ is $(\kappa,F)$-extendable. 
(It holds that $\texttt{Index}(\vec{x}(\ri(\kappa)\leftarrow \nu'))=\texttt{Index}(\vec{y}(\ri(\kappa)\leftarrow \nu'))$.
Lemma~\ref{lem:index-and-extensions} shows that if $\vec{x}(\ri(\kappa)\leftarrow\nu')$ is $(\kappa,F)$-extendable, then so is $\vec{y}(\ri(\kappa)\leftarrow\nu')$.
Strictly speaking Lemma~\ref{lem:index-and-extensions} is not applicable since it is not guaranteed that $\nu'\in\texttt{Rep}(\vec{y},\kappa,F)$ because $\nu'$ might not be a valley in the corresponding subsequence of $T$ (cf.\ Condition~\ref{tag:c2'}).
However, all other conditions are satisfied and this suffices to prove Lemma~\ref{lem:index-and-extensions}.)
Therefore, this modified array data structure is equivalent to the original data structure described in Section~\ref{subsec:algdesc}.
Thus, we have shown that Lines~\ref{line:ARA:6} to \ref{line:ARA:10} have a runtime of $\mathcal{O}(n)$, if we modify the array data structure to also allow for pointers.
This concludes our proof.
\qed\end{proof}

We conclude this section about the runtime of the alternating run algorithm by proving that an even smaller constant than $1.784$ can be expected.
Indeed, the following holds:

\begin{theorem}
Let $R_n$ be the random variable counting the number of alternating runs in an $n$-permutation chosen uniformly at random amongst all $n$-permutations.
Then for $n \geq 2$ we have: $\mathbb{E}\left(1.784^{R_n} \right) = \mathcal{O}\left(1.515^n\right)$.
\end{theorem}

\begin{proof}
In the following, let $R_{n,m}$ denote the number of $n$-permutations with exactly $m$ alternating runs.
Then the mean of $R_n$ is given as follows:
\[
\mathbb{E}(R_n)=\sum_{m \geq 1} m \cdot \frac{R_{n,m}}{n!}.
\]
By the law of the unconscious statistician (see any textbook on probability theory, e.g.\ \cite{klenke2008probability}) we then have that:
\[
\mathbb{E}\left(1.784^{R_n} \right)=\sum_{m \geq 1} 1.784^m \cdot \frac{R_{n,m}}{n!}.
\]
Let $R_n(u)=\sum_{m \geq 1} R_{n,m} u^m$ denote the generating function of alternating runs in $n$-permutations. 
Then $\mathbb{E}\left(1.784^{R_n} \right)$ can also be expressed as follows:
\[
\mathbb{E}\left(1.784^{R_n} \right)= \frac{R_{n}(1.784)}{n!}.
\]
A lot is known about the numbers $R_{n,m}$ as well as the associated generating functions: for instance $\mathbb{E}(R_n)=\frac{2n-1}{3}$ and  $\mathbb{V}(R_n)=\frac{16n-29}{90}$ (see e.g.\cite{levene1944covariance}).
However we cannot get our hands on  $R_n(1.784)$ directly, but we can do so by exploiting a connection to the well-studied Eulerian polynomials (see e.g. \cite{bona_combinatorics_2004}).
The $n$-th Eulerian polynomial $A_n(u)$ enumerates $n$-permutations by their ascents and is defined as $A_n(u)=\sum_{m \geq 1} A_{n,m} u^m$, where $A_{n,m}$ is the number of $n$-permutations with exactly $m$ ascents.
An ascent of a permutation $\pi$ is a position $i$ for which it holds that $\pi(i)< \pi(i+1)$.
Now, for the Eulerian polynomials, the following is known:
\begin{align}
\sum_{n \geq 0} A_n(u) \frac{z^n}{n!} = \frac{1-u}{e^{(u-1)z}-u}.
\label{eqn:euler_poly}
\end{align}
Moreover, we have the following connection between $R_n(u)$ and $A_n(u)$ for all integers $n \geq 2$ (established in~\cite{david1962combinatorial} and formulated more concisely by Knuth \cite{DBLP:books/aw/Knuth68}):
\[
\frac{R_n(u)}{n!}= \left(\frac{1+u}{2}\right)^{n-1} (1+w)^{n+1} A_n\left( \frac{1-w}{1+w}\right),
\]
where $w=\sqrt{(1-u)/(1+u)}$.
In order to evaluate $R_n(u)$ at $u=1.784$, we thus only need to determine $A_n(u)$ at the corresponding value.
As demonstrated in Example IX.12 in~\cite{flajolet2009analytic}, it is easy to get asymptotics for the coefficients of $z^n$  in $\sum_{n \geq 0} A_n(u) \frac{z^n}{n!}$ by a straight-forward analysis of the singularities. 
Indeed, for $|u|<2$, one has:
\begin{align}
\frac{A_n(u)}{n!}= \left( \frac{u-1}{\log(u)}\right)^{n+1} + \mathcal{O}(2^{-n}).
\label{eqn:euler_asymp}
\end{align}
Putting Equations~\eqref{eqn:euler_poly} and \eqref{eqn:euler_asymp} together, we finally obtain:
\[
\mathbb{E}\left(1.784^{R_n} \right)= \frac{R_{n}(1.784)}{n!}=\mathcal{O}\left(\left( \frac{2.784}{2}\cdot (1+w) \cdot \frac{\frac{1-w}{1+w}-1}{\log(\frac{1-w}{1+w})}\right)^n\right)=\mathcal{O}\left(c^n\right),
\]
where $w=w=\sqrt{(1-1.784)/(1+1.784)}$. Computations using any computer algebra system show that the constant $c<1.515$.
Finally, we remark that the tempting approach $\mathbb{E}\left(1.784^{R_n} \right)= 1.784^{\mathbb{E}(R_n)}$ is not correct.
\end{proof}

\begin{corollary}
The runtime of the alternating run algorithm can be expected to be in $\mathcal{O}\left(1.514^{\run(T)}\cdot n \cdot k\right)$.
\end{corollary}

\section{The parameter $\run(P)$}
\label{sec:hard}

The aim of this section is twofold:
First, we want to show that \ppm can be solved in time $\bigO(n^{1+\run(P)})$. 
This result builds upon an algorithm by Ahal and Rabinovich~\cite{DBLP:journals/siamdm/AhalR08} and a novel connection between the pathwidth of the \emph{incidence graph of a permutation}~\cite{DBLP:journals/siamdm/AhalR08} and the number of alternating runs in that permutation.
Second, we show that this runtime cannot be improved to an FPT result unless $\fpt=\w{1}$.
Let us start by defining incidence graphs:

\begin{definition}
Given an $m$-permutation $\pi$, the incidence graph $G_\pi=(V,E)$ of $\pi$ is defined as follows:
The vertices $V\colonequals[m]$ represent positions in $\pi$.
There are edges between adjacent positions, i.e., $E_1\colonequals\big\{\{i,i+1\}\mid i\in[m-1]\big\}$.
There are also edges between positions where the corresponding elements have a difference of $1$, i.e., $E_2\colonequals\big\{\{i,j\}\mid \pi(i)-\pi(j)=1\big\}$.
The edge set is defined as $E\colonequals E_1\cup E_2$.
\end{definition}

\begin{example}
\label{ex:pathwidth}
Consider the permutation 
\[
\pi = \left( 
\begin{tabular}{ccccccccc}
1 & 2 & 3 & 4 & 5 & 6 & 7 & 8 & 9 \\ 
2 & 5 & 9 & 7 & 4 & 6 & 8 & 3 & 1 \\ 
\end{tabular} 
\right) 
\]
written in two-line representation. 
A graphical representation of $\pi$ can be found on the left-hand side of Figure~\ref{fig:ex_pathwidth}. 
The corresponding graph $G_\pi$ is represented on the right-hand side of the same figure.
The solid lines correspond to the edges in $E_1$ and the dashed lines to the ones in $E_2$.
\demo
\begin{figure}
\hspace{1cm}
\begin{minipage}{0.47\textwidth}
\label{fig:example_G1}
\begin{tikzpicture}[thick, scale=0.62]
	\tikzstyle{every node}=[rectangle,  draw=black,  text centered,  font=\normalsize] 
	
  	\foreach \x/\y in {1/2, 2/5, 3/9, 4/7, 5/4, 6/6, 7/8, 8/3, 9/1}
    \node (t\x) at (\x, \y) {\y}; 
    ;
  	
	\foreach \x/\y in {1/2,2/3,3/4,4/5,5/6,6/7,7/8,8/9}
			\draw (t\x) -- (t\y);
\end{tikzpicture}
\end{minipage}
\begin{minipage}{0.6\textwidth}
\end{minipage}
\begin{minipage}{0.47\textwidth}
\begin{tikzpicture}[thick, scale=0.62]
	\tikzstyle{every node}=[circle,  draw=black,  text centered, font=\normalsize] 
	
  	\foreach \x/\y in {1/2, 2/5, 3/9, 4/7, 5/4, 6/6, 7/8, 8/3, 9/1}
    \node (t\x) at (\x, \y) {\x}; 
    ;
  	
	\foreach \x/\y in {1/2,2/3,3/4,4/5,5/6,6/7,7/8,8/9}
			\draw (t\x) -- (t\y);
	\foreach \x/\y in {9/1,1/8,8/5,5/2,2/6,6/4,4/7,7/3}
			\draw[dashed] (t\x) -- (t\y);
\end{tikzpicture}
\end{minipage}
\caption{To the left is a graphical representation of the permutation $\pi$ introduced in Example~\ref{ex:pathwidth}, to the right the corresponding incidence graph $G_\pi$.}
\label{fig:ex_pathwidth}
\end{figure}
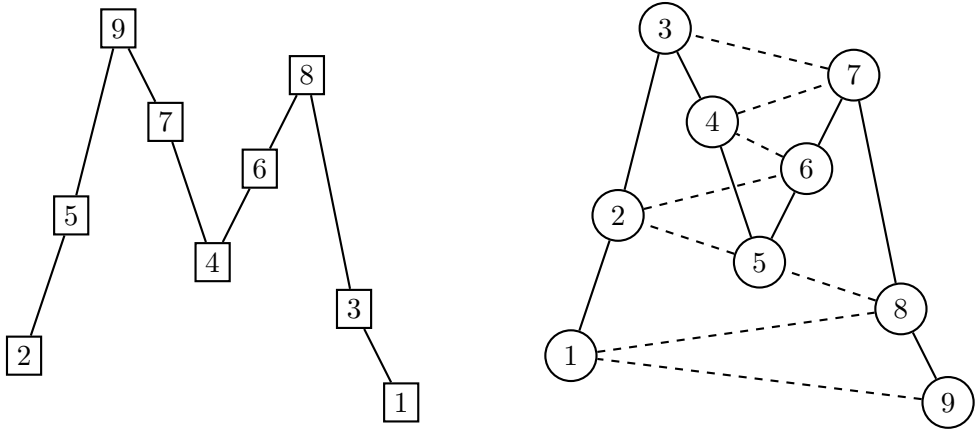
\end{example}

\begin{definition}
Let $G=(V,E)$ be a simple graph, i.e., $E$ is a set of cardinality~$2$ subsets of $V$.
A path decomposition of $G$ is a sequence of subsets $S_i\subseteq V$ such that
\begin{enumerate}
\item Every vertex appears in at least one $S_i$.
\item Every edge is a subset of at least one $S_i$.
\item Let three indices $h<i<j$ be given. If a vertex is contained both in $S_h$ and $S_j$ then it is also contained in $S_i$.
\end{enumerate}
The width of a path decomposition is defined as $\max_i\card{S_i}-1$.
The pathwidth of a graph $G$, written $\mathsf{pw}(G)$, is the minimum width of any path decomposition.
\end{definition}

In \cite{DBLP:journals/siamdm/AhalR08}, Theorem 2.7 and Proposition 3.5, the authors present an algorithm that solves \ppm in time $\bigO\left(n^{1+\mathsf{pw}(G_P)}\right)$.
The following lemma relates $\mathsf{pw}(G_P)$ and the number of alternating runs in $P$.

\begin{lemma}
\label{lem:xp}
For all permutations $\pi$, it holds that $\mathsf{pw}(G_\pi)\leq \run(\pi)$.
\end{lemma}

\begin{proof}
Given an $m$-permutation $\pi$ we will define a sequence $S_1,\ldots,S_m$.
We then show that this sequence is a path decomposition of $G_\pi=(V,E)$ with width at most $\run(\pi)$.

In order to define the sequence $S_1,\ldots,S_m$ of subsets of $V$, we shall extend alternating runs to maximal monotone subsequences. 
This means that we add the preceding valley to a run up and the preceding peak to a run down.
For any $i\in[\run(\pi)]$, $R_i$ then denotes the set of elements in the $i$-th run in $\pi$ together with the preceding valley or peak.
Note that this implies that $|R_i \cap R_{i+1}|=1$ for all $i\in[\run(\pi)-1]$.

We define $S'_1\colonequals\{1\}$ and for every $v\in[2,m]$, \[S'_v\colonequals \big\{ \max(R_j\cap[v-1])\mid j\in[\run(\pi)]\text{ and } R_j\cap[v-1]\neq\emptyset\big\}\cup\big\{v\big\},\]
i.e., $S_v'$ contains $v$ and the largest element of every run that is smaller than $v$.
Since $S_v$ should contain positions in $\pi$ (and not elements), we define \[S_v\colonequals\{\pi^{-1}(w)\mid w\in S'_v\}.\]

For an example of this construction, see Example~\ref{ex:pw,run}.
We now check that $S_1,\ldots,S_m$  indeed is a path decomposition.
\begin{enumerate}
\item The vertex $i$ appears in $S_{\pi(i)}$.
\item First we consider edges of the form $\{i,i+1\}$.
Without loss of generality let $\pi(i)<\pi(i+1)$.
Then $\{i,i+1\}$ is a subset of $S_{\pi(i+1)}$.
Clearly, $i+1\in S_{\pi(i+1)}$.
Since $\pi(i)$ and $\pi(i+1)$ are adjacent in $\pi$ there has to be an $s\in[\run(\pi)]$ such that $\{\pi(i),\pi(i+1)\}\subseteq R_s$.
It then holds that $\max(R_s\cap[\pi(i+1)-1])=\pi(i)$ since $\pi(i) \in R_s\cap[\pi(i+1)-1]$ and $\pi(i)$ is the largest element in $R_s$ smaller than $\pi(i+1)$.
Consequently $i\in S_{\pi(i+1)}$.

Second, every edge $\{i,j\}\in E$ with $\pi(i)-\pi(j)=1$ is a subset of $S_{\pi(i)}$:
As before $i\in S_{\pi(i)}$.
Let $s$ be any element of $[\run(\pi)]$ such that $j\in R_s$.
Then $\max(R_s\cap[\pi(i)-1])=\max(R_s\cap[\pi(j)])=\pi(j)$ and hence $j\in S_{\pi(i)}$.

Only these two types of edges exists.
\item Let $1 \leq u<v<w\leq m$ with $i\in S_u$ and $i\in S_w$.
Let $s$ be any element of $[\run(\pi)]$ such that $\pi(i)\in R_s$.
Then either $\pi(i)\in R_s\cap[u-1]$ or $\pi(i)=u$.
In both cases is $\pi(i) \in R_s\cap[v]$.
Furthermore, since $\pi(i)<w$, $\pi(i)=\max(R_s\cap[w-1])=\max(R_s\cap[v])$.
Hence $\pi(i)\in S'_v$ and $i\in S_v$.
\end{enumerate}
The cardinality of each $S_i$ is at most $\run(\pi)+1$ and hence $\mathsf{pw}(G_\pi)\leq\run(\pi)$.
\qed\end{proof}

\begin{remark}
This bound is tight: for $\pi=\perm{1,2,3,\ldots,m}$ the graph $G_\pi$ is a path and hence $\mathsf{pw}(G_\pi)=\run(\pi)=1$.
\end{remark}

\begin{example}
Consider again $\pi$ as defined in Example~\ref{ex:pathwidth}. 
The elements of the sets $S'_1, \ldots, S'_9$ and those of $S_1, \ldots, S_9$ as defined in the proof of Lemma~\ref{lem:xp} are given in Figure~\ref{tab:pathwidth}. 
It is easy to check that $S_1, \ldots, S_9$ indeed is a path decomposition of width $4=\run(\pi)$.
Note that in the given table, columns of equal numbers do not contain any gaps.
This fact corresponds to the third condition in the definition of path decompositions. 
\label{ex:pw,run}
\demo
\begin{figure}
\label{tab:pathwidth}
\begin{center}
\begin{tabular}{|p{0.25cm}|p{1.9cm}|p{1.9cm}|}
\hline 
$v$ &  $S'_v$ & $S_v$ \\ 
\hline 
1  & \texttt{1} & \texttt{9} \\ 
\hline 
2  & \texttt{12} & \texttt{91} \\ 
\hline 
3  & \texttt{123} & \texttt{918} \\ 
\hline 
4  & \texttt{ 234} & \texttt{\ 185} \\ 
\hline 
5  & \texttt{ 2345} & \texttt{\ 1852} \\ 
\hline 
6  & \texttt{ \ 3456} & \texttt{\ \ 8526} \\ 
\hline 
7  & \texttt{ \ 34567} & \texttt{\ \ 85264} \\ 
\hline 
8  & \texttt{ \ 3 5678} & \texttt{\ \ 8\ 2647} \\ 
\hline 
9  & \texttt{\ \ \ \ 5\ 789} & \texttt{\ \ \ \ 2\ 473} \\ 
\hline 
\end{tabular}
\end{center}
\caption{The sets $S'_1, \ldots, S'_9$ and $S_1, \ldots, S_9$ for the permutation $\pi=2\,5\,9\,7\,4\,6\,8\,3\,1$}
\end{figure}
\end{example}

\begin{theorem}
\label{thm:xp}
\ppm can be solved in time $\bigO(n^{1+run(P)})$.
\end{theorem}
\begin{proof}
Since $\mathsf{pw}(G_\pi)\leq \run(\pi)$ (Lemma~\ref{lem:xp}), the runtime of the $\bigO(n^{1+\mathit{pw}(G_P)})$ algorithm can be bounded by $\bigO(n^{1+run(P)})$.\qed
\end{proof}

We continue with a corresponding hardness result.
We prove that one cannot hope to substantially improve the \xp results in Theorem~\ref{thm:xp}:
an FPT algorithm with respect to $\run(P)$ is only possible if $\fpt=\w{1}$.

\begin{theorem}
\ppm is \w{1}-hard with respect to the parameter $\run(P)$.
\label{thm:w1_completeness_run(P)}
\end{theorem}

\begin{proof} 
We give an FPT-reduction from the \w{1}-hard \textsc{Segregated Permutation Pattern Matching Problem}~\cite{landscape} to \ppm. 
This problem is defined as follows:

\pprob{\textsc{Segregated Permutation Pattern Matching (SPPM)}}
{An $n$-\pea $T$ (the text), a $k$-permutation $P$ (the pattern) and two positive integers $p\in[k]$, $t\in[n]$.}
{$k$}
{Is there a matching $M$ of $P$ into $T$ such that $M(u)\leq t$ if and only if $u\leq p$?}

In this problem we are looking for matchings $M$ where for all $u\leq p$ it holds that $M(u)\in [t]$ and for all $u> p$ it holds that $M(u)\in [t+1,n]$.
Let $(P,T,p,t)$ be a \textsc{SPPM} instance, where $|P|=k\leq n=|T|$.
We are going to construct a \ppm instance $(\tilde{P}, \tilde{T})$ as follows:
\begin{align*}
\tilde{P} & = (p+0.5) \quad \underbrace{(k+1)(k+2) \ldots (k+n+1)}_{=R_P}  \quad P \\
\tilde{T} & = (t+0.5) \quad \underbrace{(n+1)(n+2) \ldots \ldots (2n+1)}_{=R_T} \quad T
\end{align*} 
Note that the increasing runs $R_P$ and $R_T$ both consist of $(n+1)$ elements.
The element placed at the beginning of $\tilde{P}$, $p+0.5$, is larger than $p$ but smaller than $p+1$.
Analogously, $t+0.5$ in $\tilde{T}$ is larger than $t$ but smaller than $t+1$. 
Note that $\tilde{P}$ and $\tilde{T}$ are not permutations in the classical sense, since they contain elements that are not integers.
However, in order to obtain permutations on $[k+n+2]$ and $[2n+2]$, we simply need to relabel the respective elements order-isomorphically.

Given this construction of $\tilde{P}$ and $\tilde{T}$ the following holds: In every matching of $\tilde{P}$ into $\tilde{T}$ the element $p+0.5$ has to be mapped to $t+0.5$.
Indeed, the increasing run of elements $R_P=(k+1)(k+2) \ldots (k+n+1)$ in $\tilde{P}$ has to be mapped to the increasing run of elements $R_T=(n+1)(n+2) \ldots \ldots (2n+1)$ in $\tilde{T}$ and consequently $P$ has to be matched into $T$.
This holds because of the following observation: 
If the element $(k+1)$ in $\tilde{P}$ is mapped to an element $(n+u)$ with $u >1$ in $\tilde{T}$, some of the elements of $R_P$ have to be matched into $T$ since $R_P$ and $R_T$ have the same length. 
This is however not possible, since all elements in $T$ are smaller than $(n+u)$.
If $(k+1)$ is instead mapped to one of the elements of $T$, then all remaining elements of $R_P$ also have to be matched into $T$ which is not possible since $R_P$ is longer than $T$.
Therefore, the element $(k+1)$ in $\tilde{P}$ is always mapped to the element $(n+1)$ in $\tilde{T}$.
Both in $\tilde{P}$ and in $\tilde{T}$ there is only one element lying to the left of $(k+1)$ and one to left of $(n+1)$: $(p+0.5)$ and $(t+0.5)$, respectively.
Thus, $(p+0.5)$ has to be mapped to $(t+0.5)$. 
This implies that all elements smaller than $(p+0.5)$, i.e., elements in the interval $[p]$, in $P$ have to be mapped to elements smaller than $t+0.5$, i.e., elements in the interval $[t]$, in $T$.
We have shown that $(P,T,p,t)$ is a YES-instance of \textsc{SPPM} if and only if $(\tilde{P}, \tilde{T})$ is a YES-instance of \ppm.

It remains to show that this reduction can be done in FPT time. 
Clearly $\run(\tilde{P})=2+\run(P)=\mathcal{O}(k)$.
Moreover the length of $T$ is bounded by a polynomial in the size of $G$ since $|T|=n+2+|T|=2n+2=\mathcal{O}(n)$.
\qed\end{proof}

\section{Research directions}
\label{sec:future}
Theorem \ref{thm-ppm-altruns} shows fixed-parameter tractability of \ppm with respect to $\run(T)$. 
An immediate consequence is that any \ppm instance can be reduced by polynomial time preprocessing to an equivalent instance -- a kernel -- of size depending solely on $\run(T)$. 
This raises the question whether even a polynomial-sized kernel exists.
Such kernels, and in particular polynomial kernels, have been the focus of intensive research in algorithmics~\cite{invitationtokernelization}.
Another research direction is the study of further parameters such as permutation statistics listed in the Appendix A of~\cite{kitaev2011patterns}.

At this point, several algorithms exist that solve \ppm without imposing restrictions on $P$ and $T$.
The algorithms by Guillemot and Marx~\cite{guillemotmarx2013ppmfpt}, Albert et al.~\cite{springer:AlbertAAH01} and Ahal and Rabinovich~\cite{DBLP:journals/siamdm/AhalR08} seem to be particularly well-suited for small patterns.
In contrast, the runtime of our algorithm does not depend that critically on $\card{P}$.
Thus, it may be expected that our algorithm is preferable for large patterns.
However, only implementations and benchmarks could allow to settle this question and compare these algorithms.

Finally, our method of making use of alternating runs might also lead to fast algorithms for other permutation based problems.

\section{Acknowledgments}
\label{sec:ack}
We would like to thank the anonymous reviewers for their feedback and suggestions leading to numerous improvements of this paper.
The first author was supported by the Austrian Science Foundation FWF, grant P25337-N23, the second author by the Austrian Science Foundation FWF, grant P25518-N23 and Y698.

\bibliographystyle{spmpsci}
\bibliography{lit}

\end{document}